\documentclass[12pt]{article}
\usepackage{amsfonts,graphicx,amsmath,amsthm,amssymb,epsfig}
\usepackage{float}
\allowdisplaybreaks
\begin{document}

\title{\bf Decoupled Anisotropic Buchdahl's Relativistic Models in $f(\mathbb{R},\mathbb{T})$ Theory}
\author{Tayyab Naseer \thanks{tayyabnaseer48@yahoo.com;
tayyab.naseer@math.uol.edu.pk}~ and M. Sharif \thanks{msharif.math@pu.edu.pk} \\
Department of Mathematics and Statistics, The University of Lahore,\\
1-KM Defence Road Lahore-54000, Pakistan.}

\date{}
\maketitle

\begin{abstract}
This paper constructs three different anisotropic extensions of the
existing isotropic solution to the modified field equations through
the gravitational decoupling in $f(\mathbb{R},\mathbb{T})$ theory.
For this, we take a static sphere that is initially filled with the
isotropic fluid and then add a new gravitational source producing
anisotropy in the system. The field equations now correspond to the
total matter configuration. We transform the radial metric component
to split these equations into two sets characterizing their parent
sources. The unknowns comprising in the first set are determined by
considering the Buchdahl isotropic solution. On the other hand, we
employ different constraints related to the additional gravitational
source and make the second system solvable. Further, the constant
triplet in Buchdahl solution is calculated by means of matching
criteria between the interior and exterior geometries at the
spherical boundary. The mass and radius of a compact star LMC X-4
are used to analyze the physical relevancy of the developed models.
We conclude that our resulting models II and III are in
well-agreement with acceptability conditions for the considered
values of the parameters.
\end{abstract}
{\bf Keywords:} Modified theory;
Anisotropy; Gravitational decoupling. \\
{\bf PACS:} 04.40.Dg; 04.50.Kd; 04.40.-b.

\section{Introduction}

Cosmologists have recently made significant discoveries that
challenge the perception of the arrangement of astrophysical
structures in our universe. Rather than a random distribution, these
structures appear to be systematically organized, sparking
considerable interest among researchers. The detailed study of such
interstellar bodies has become a focal point for scientists seeking
to figure out the mystery of the accelerated expansion of the
cosmos. It has become evident through various experiments that our
universe contains abundance of a force countering the pull of
gravity, and thus driving such expansion. This mysterious force is
referred to the dark energy due to its obscure nature that continues
to confuse researchers. While general relativity ($\mathbb{GR}$)
explains this expansion up to some extent, it grapples with issues
related to the cosmological constant. Therefore, some extensions of
this theory need to be proposed.

A straightforward generalization to $\mathbb{GR}$ is the
$f(\mathbb{R})$ theory that represents a significant advancement in
the field of theoretical physics. It introduces a modification to
the Einstein-Hilbert action, where the roles of the Ricci scalar
$\mathbb{R}$ and its generic function are interchanged. This theory
has provided promising results and has been employed to investigate
self-gravitating systems \cite{2}-\cite{9}. Astashenok et al.
\cite{1n} investigated the causal limit of maximum mass for stars in
this modified framework. Their findings led to the conclusion that
the secondary component of the compact binary GW190814 is likely to
be a neutron star, a black hole, or possibly a rapidly rotating
neutron star, ruling out the possibility of it being a strange star.
There is a body of literature, pointing out such important works by
various researchers \cite{1o}-\cite{1t}. Bertolami et al. \cite{10}
initially presented the idea of studying the fluid-geometry
interaction in $f(\mathbb{R})$ framework and executed this by
combining the matter Lagrangian density and $\mathbb{R}$ as a single
function. This notion prompted astronomers to put their focus on the
discussion of the rapid expansion of the universe \cite{11}.

Soon after this, Harko et al. \cite{20} generalized this concept at
the action level, and pioneered a new gravitational theory. This was
named as $f(\mathbb{R},\mathbb{T})$ theory where the effects of
geometry and matter configuration are coupled through $\mathbb{R}$
and trace of the energy-momentum tensor ($\mathbb{EMT}$)
$\mathbb{T}$. This generalized function results in the non-conserved
system, hence, an extra force (that depends on physical parameters
like pressure and density \cite{22}) appears in the gravitational
field which forces the test particles to move in a non-geodesic
path. Houndjo \cite{22a} employed a minimal model of this modified
theory to explain the conversion of the matter-dominated phase into
late-time acceleration era successfully. Among several
$f(\mathbb{R},\mathbb{T})$ models, $\mathbb{R}+2\varpi\mathbb{T}$
gained much importance in the literature that produces physically
acceptable compact interiors. This model is adopted by Das et al.
\cite{22b} to establish a gravastar-like model consisting of three
layers, each of them is represented by different equations of state.
Multiple stellar interiors have been discussed by using various
approaches in the context of this modified model
\cite{25ae}-\cite{25adb}. One significant notion of
$f(\mathbb{R},\mathbb{T})$ theory is that it considers the quantum
effect, leading the possibilities of particle production. This
aspect holds much importance in astrophysical research as it posits
a connection between quantum theory and $f(\mathbb{R},\mathbb{T})$
gravity. Some notewrothy applications of this theory in both
astrophysics  and cosmology can be seen in \cite{1i}-\cite{1k}.
Recently, Zaregonbadi et al. \cite{1l} have examined the viability
of this modified theory to explore the effects of dark matter on the
galactic scale.

The standard model of the cosmos is believed to be mainly based on
the homogeneity and isotropy at a large-scale, however, there exists
pressure anisotropy at some small-scales \cite{25ag}-\cite{25ah}.
Further, the geometry of compact objects guarantees the presence of
anisotropic pressure and inhomogeneous fluid distribution. The
former factor appears when there is a difference between the
pressure in both radial and transverse directions. There is a class
of components that produces anisotropy in the interior configuration
such as phase transitions \cite{25ai}, pion condensation
\cite{25aj}, neutron stars surrounded by a strong electromagnetic
field \cite{25ak}, and some other factors \cite{25al,25ala}. Another
factor that causes the anisotropy is the gravitational effects
produced by the tidal forces \cite{25am}. The isotropy of our
universe has recently been examined by analyzing X-rays coming from
galactic clusters \cite{25an}. The same strategy has then been
carried out on massive structures, and it was concluded that the
nature of cosmos is anisotropic \cite{25ao}. The pressure anisotropy
is thus much significant to be studied and affects several physical
characteristics like the energy density, gravitational redshift and
mass, etc., of a self-gravitating system.

The self-gravitating celestial objects represented by non-linear
gravitational field equations in $\mathbb{GR}$ or other modified
theories prompted astrophysicists to obtain their exact/numerical
solutions. The compact interiors could be of physical interest only
if the corresponding formulated solution fulfills the required
conditions. Multiple techniques, in this regard, have recently been
suggested, one of them is the gravitational decoupling that is
employed to model a compact interior possessing multiple sources
including anisotropy, heat dissipation, shear, etc. The initial idea
of this technique was based on the fact that the field equations
comprising different sources can be decoupled into multiple sets,
and thus it becomes an easy task to solve each set individually.
Ovalle \cite{29} recently pioneered the minimal geometric
deformation (MGD) scheme which provides some enticing ingredients to
formulate physically acceptable stellar solutions in the braneworld.
Following this, an isotropic spherically symmetric matter
distribution was discussed by Ovalle and Linares \cite{30} who
formulated an analytical solution in the braneworld and found it
consistent with Tolman-IV ansatz. Casadio et al. \cite{31} obtained
the Schwarzschild geometry in the context of Randall-Sundrum
braneworld theory by extending the above strategy.

Ovalle and his collaborators \cite{33} adopted a spherically
symmetric isotropic interior and developed its physically feasible
anisotropic version using the MGD scheme. Sharif and Sadiq \cite{34}
introduced the straightforward extension of this technique to the
charged case where they constructed two different anisotropic
counterparts of Krori-Barua metric potentials and discussed their
stability. Multiple metric ansatz have been extended to obtain their
corresponding anisotropic analogs through MGD in modified
$f(\mathbb{R})$ theory \cite{35}. The Durgapal-Fuloria metric
coefficients have been taken as a seed isotropic source through
which several physically relevant anisotropic solutions were
obtained \cite{36}. Different researchers proposed the anisotropic
extensions of isotropic Heintzmann as well as Tolman VII solutions
and found them stable in the considered range \cite{36a,37a}. Sharif
and Ama-Tul-Mughani \cite{37b} took the axial spacetime into account
and formulated their corresponding well-behaved solutions. We have
also obtained such charged/uncharged anisotropic analogs of
Krori-Barua ansatz in a strong non-minimally coupled gravitational
theory \cite{37f}-\cite{37fc}.

This article formulates three different anisotropic solutions that
are, in fact, the extensions of an isotropic interior to the
additional matter source coupled gravitationally to the seed source
in $f(\mathbb{R},\mathbb{T})$ framework. The following lines present
how this paper is organized. The fundamentals of the modified
$f(\mathbb{R},\mathbb{T})$ theory and its relevant field equations
corresponding to the total (seed and newly added) matter source are
formulated in the next section. Section \textbf{3} introduces the
MGD transformation that helps to decouple the field equations into
two sets. We then adopt the Buchdahl's solution in section
\textbf{4} and calculate the unknown constants through boundary
conditions. Section \textbf{5} presents some conditions whose
fulfillment leads to physically acceptable model. Three newly
developed anisotropic solutions and their graphical interpretation
are provided in section \textbf{6}. Lastly, we summarize our
outcomes in the last section.

\section{$f(\mathbb{R},\mathbb{T})$ Gravity}

The Einstein-Hilbert action for the modified
$f(\mathbb{R},\mathbb{T})$ theory becomes (with $\kappa=8\pi$) after
the inclusion of an additional field as \cite{20}
\begin{equation}\label{g1}
S=\int \sqrt{-g}\left[\frac{f(\mathbb{R},\mathbb{T})}{16\pi}
+\pounds_{m}+\eta\pounds_{\mathfrak{D}}\right]d^{4}x,
\end{equation}
where $\pounds_{m}$ is the fluid's Lagrangian density. We suppose an
extra source to be gravitationally coupled with the parent matter
configuration whose corresponding Lagrangian is denoted by
$\pounds_{\mathfrak{D}}$. The decoupling parameter $\eta$ explains
how much an extra source influences the physical properties
characterizing a self-gravitating system. Also, the metric tensor
$g_{\sigma\omega}$ in this case provides the determinant indicated
by $g$. The least action principle is applied on the action
\eqref{g1}, leading to the field equations given by
\begin{equation}\label{g2}
\mathbb{G}_{\sigma\omega}=8\pi
\mathbb{T}_{\sigma\omega}^{\mathrm{(tot)}},
\end{equation}
where $\mathbb{G}_{\sigma\omega}$ and
$\mathbb{T}_{\sigma\omega}^{\mathrm{(tot)}}$ describe the geometric
sector and the interior fluid distribution, respectively. The later
term is further classified as
\begin{equation}\label{g3}
\mathbb{T}_{\sigma\omega}^{\mathrm{(tot)}}=\mathbb{T}_{\sigma\omega}^{\mathrm{(eff)}}+\eta
\mathfrak{D}_{\sigma\omega}=\frac{1}{f_{\mathbb{R}}}\mathbb{T}_{\sigma\omega}+\mathbb{T}_{\sigma\omega}^{(C)}+\eta
\mathfrak{D}_{\sigma\omega}.
\end{equation}
Here $\mathfrak{D}_{\sigma\omega}$ is the additional fluid source.
Moreover, the effective matter sector
$\mathbb{T}_{\sigma\omega}^{\mathrm{(eff)}}$ is divided into the
usual and modified $\mathbb{EMT}s$. The later term
$\mathbb{T}_{\sigma\omega}^{(C)}$ takes the form
\begin{eqnarray}
\nonumber \mathbb{T}_{\sigma\omega}^{(C)}&=&\frac{1}{8\pi
f_{\mathbb{R}}}\bigg[f_{\mathbb{T}}\mathbb{T}_{\sigma\omega}+\bigg\{\frac{\mathbb{R}}{2}\bigg(\frac{f}{\mathbb{R}}
-f_{\mathbb{R}}\bigg)-\pounds_{m}f_{\mathbb{T}}\bigg\}g_{\sigma\omega}\\\label{g4}
&-&(g_{\sigma\omega}\Box-\nabla_{\sigma}\nabla_{\omega})f_{\mathbb{R}}+2f_{\mathbb{T}}g^{\zeta\beta}\frac{\partial^2
\pounds_{m}}{\partial g^{\sigma\omega}\partial
g^{\zeta\beta}}\bigg],
\end{eqnarray}
where $f_{\mathbb{T}}$ and $f_{\mathbb{R}}$ mean $\frac{\partial
f(\mathbb{R},\mathbb{T})}{\partial \mathbb{T}}$ and $\frac{\partial
f(\mathbb{R},\mathbb{T})}{\partial \mathbb{R}}$, respectively.
Furthermore,
$\Box\equiv\frac{1}{\sqrt{-g}}\partial_\sigma\big(\sqrt{-g}g^{\sigma\omega}\partial_{\omega}\big)$
is the D'Alembert operator and $\nabla_\sigma$ is the covariant
divergence. Also, we adopt $\pounds_{m}=P$ ($P$ is an isotropic
pressure), leading to $\frac{\partial^2 \pounds_{m}}{\partial
g^{\sigma\omega}\partial g^{\zeta\beta}}=0$.

We consider that the geometrical structure is initially filled with
an isotropic fluid which can be expressed through the following
$\mathbb{EMT}$
\begin{equation}\label{g5}
\mathbb{T}_{\sigma\omega}=(\mu+P)\mathcal{K}_{\sigma}\mathcal{K}_{\omega}+P
g_{\sigma\omega},
\end{equation}
where $\mu$ and $\mathcal{K}_{\sigma}$ are the energy density and
four-velocity, respectively. We obtain the trace of Eq.\eqref{g2} as
follows
\begin{align}\nonumber
&2f+\mathbb{T}(f_\mathbb{T}+1)-\mathbb{R}f_\mathbb{R}-3\nabla^{\sigma}\nabla_{\sigma}f_\mathbb{R}-4f_\mathbb{T}\pounds_m
+2f_\mathbb{T}g^{\zeta\beta}g^{\sigma\omega}\frac{\partial^2\pounds_m}{\partial
g^{\zeta\beta}\partial g^{\sigma\omega}}=0.
\end{align}
Since the action of this theory involves matter-geometry coupled
functional, the disappearance of the matter term, i.e.,
$\mathbb{T}=0$ (or vacuum case) reduces all our results in
$f(\mathbb{R})$ gravity. Moreover, the non-zero divergence of the
usual $\mathbb{EMT}$ is observed in this theory due to such strong
interaction. Resultantly, an additional force appears in the field
of self-gravitating object that alters the geodesic path of the
moving test particles. Its mathematical expression is given by
\begin{align}\nonumber
\nabla^\sigma\mathbb{T}_{\sigma\omega}&=\frac{f_\mathbb{T}}{8\pi-f_\mathbb{T}}\bigg[(\mathbb{T}_{\sigma\omega}
+\Upsilon_{\sigma\omega})\nabla^\sigma\ln{f_\mathbb{T}}+\nabla^\sigma\Upsilon_{\sigma\omega}\\\label{g11}
&-\frac{8\pi\eta}{f_\mathbb{T}}\nabla^\sigma\mathfrak{D}_{\sigma\omega}
-\frac{1}{2}g_{\zeta\beta}\nabla_\omega\mathbb{T}^{\zeta\beta}\bigg],
\end{align}
where
$\Upsilon_{\sigma\omega}=g_{\sigma\omega}\pounds_m-2\mathbb{T}_{\sigma\omega}-2g^{\zeta\beta}\frac{\partial^2
\pounds_{m}}{\partial g^{\sigma\omega}\partial g^{\zeta\beta}}$.

We consider a spherical interior geometry that is distinguished from
the exterior region at the hypersurface $\Sigma$ given by the
following line element
\begin{equation}\label{g6}
ds^2=-e^{\nu_1} dt^2+e^{\nu_2} dr^2+r^2\big(d\theta^2+\sin^2\theta
d\vartheta^2\big),
\end{equation}
where $\nu_1=\nu_1(r)$ and $\nu_2=\nu_2(r)$. The corresponding
four-velocity now takes the form in terms of temporal metric
component as
\begin{equation}\label{g7}
\mathcal{K}_\sigma=-\delta^0_\sigma
e^{\frac{\nu_1}{2}}=(-e^{\frac{\nu_1}{2}},0,0,0).
\end{equation}
In order to have some meaningful results, we adopt a standard
modified model. Although the literature presents several
matter-geometry coupled (minimal as well as non-minimal)
$f(\mathbb{R},\mathbb{T})$ models, however, we adopt a minimal one
which is given by
\begin{equation}\label{g7a}
f(\mathbb{R},\mathbb{T})=f_1(\mathbb{R})+
f_2(\mathbb{T})=\mathbb{R}+2\nu_3\mathbb{T},
\end{equation}
where $\mathbb{T}=-\mu+3P$ and $\nu_3$ is a real-valued constant.
There are two main reasons behind this choice highlighted as
\begin{itemize}
\item For complicated $f(\mathbb{R},\mathbb{T})$ functionals containing exponential, logarithmic,
or polynomials terms of $\mathbb{T}$, or higher-order curvature
terms such as $\mathbb{R}^2$, we will obtain a very complicated form
of the field equations. So, the problem appears when it comes to
split those equations into two sets using MGD technique (following
section elaborates MGD methodology thoroughly). In such cases, we
may be unable to obtain two new sectors characterizing their parent
fluid sources.

\item On the other hand, in the case of complex
$f(\mathbb{R},\mathbb{T})$ functionals, an issue appears when we
deal with matching conditions across the spherical interface.
Therefore, in order to adjust these issues, we choose the functional
given in Eq.\eqref{g7a}.
\end{itemize}
Moreover, Ashmita et al. \cite{38} derived potential slow-roll
parameters by adopting multiple inflation potentials in this
gravitational theory and found their results to be consistent with
the experimental data only when $-0.37<\nu_3<1.483$. This model has
also been used to formulate physically relevant anisotropic version
of the interior Tolman-Kuchowicz spacetime \cite{39}. We have
developed acceptable decoupled solutions corresponding to
anisotropic configuration in this theory \cite{40,41}.

The spherical spacetime \eqref{g6} produces the field equations
corresponding to the modified model \eqref{g7a} as
\begin{align}\label{g8}
&e^{-\nu_2}\left(\frac{\nu_2'}{r}-\frac{1}{r^2}\right)
+\frac{1}{r^2}=8\pi\left(\mu-\eta\mathfrak{D}_{0}^{0}\right)+\nu_3\left(3\mu-P\right),\\\label{g9}
&e^{-\nu_2}\left(\frac{1}{r^2}+\frac{\nu_1'}{r}\right)
-\frac{1}{r^2}=8\pi\left(P+\eta\mathfrak{D}_{1}^{1}\right)-\nu_3\left(\mu-3P\right),
\\\label{g10}
&\frac{e^{-\nu_2}}{4}\left[\nu_1'^2-\nu_2'\nu_1'+2\nu_1''-\frac{2\nu_2'}{r}+\frac{2\nu_1'}{r}\right]
=8\pi\left(P+\eta\mathfrak{D}_{2}^{2}\right)-\nu_3\left(\mu-3P\right),
\end{align}
where the entities along with $\nu_3$ appear as the
$f(\mathbb{R},\mathbb{T})$ corrections and prime means
$\frac{\partial}{\partial r}$. Moreover, we obtain the generalized
form of Tolman-Opphenheimer-Volkoff equation from \eqref{g11}
corresponding to the model \eqref{g7a} as
\begin{align}\nonumber
&\frac{dP}{dr}+\frac{\nu_1'}{2}\left(\mu+P\right)+\frac{\eta\nu_1'}{2}
\left(\mathfrak{D}_{1}^{1}-\mathfrak{D}_{0}^{0}\right)+\eta\frac{d\mathfrak{D}_{1}^{1}}{dr}\\\label{g12}
&+\frac{2\eta}{r}\left(\mathfrak{D}_{1}^{1}-\mathfrak{D}_{2}^{2}\right)=-\frac{\nu_3}{4\pi-\nu_3}\big(\mu'-P'\big),
\end{align}
which verifies the nature of this extended theory of gravity to be
non-conserved. Since Eq.\eqref{g12} is a combination of different
forces that maintain the hydrostatic equilibrium inside a celestial
object, it plays an important role to study the interior's
structural changes. It is observed that the extraction of the
solution of a system \eqref{g8}-\eqref{g10} becomes obscure due to
the entanglement of a large number of unknowns, i.e.,
$(\nu_1,\nu_2,\mu,P,\mathfrak{D}_{0}^{0},\mathfrak{D}_{1}^{1},\mathfrak{D}_{2}^{2})$.
Therefore, it becomes necessary to adopt some constraints,
otherwise, the system cannot be solved uniquely. In this context, a
systematic scheme \cite{33} is adopted to fulfill our requirement.

\section{Minimal Gravitational Decoupling}

Gravitational decoupling is an efficient approach that transforms
the metric potentials in a new reference frame and makes it easy to
construct the solution of highly non-linear field equations
representing compact systems. For this, a new line element is
considered as a solution to the field equations
\eqref{g8}-\eqref{g10} given by
\begin{equation}\label{g15}
ds^2=-e^{\nu_4(r)}dt^2+\frac{1}{\nu_5(r)}dr^2+r^2\big(d\theta^2+\sin^2\theta
d\vartheta^2\big).
\end{equation}
Following equations transform the metric potentials and decouple the
field equations as
\begin{equation}\label{g16}
\nu_4\rightarrow\nu_1=\nu_4+\eta\mathrm{t}_1, \quad \nu_5\rightarrow
e^{-\nu_2}=\nu_5+\eta\mathrm{t}_2,
\end{equation}
where $\mathrm{t}_1$ and $\mathrm{t}_2$ correspond to $g_{tt}$ and
$g_{rr}$ components, respectively. Gravitational decoupling offers
two different techniques, i.e., minimal and extended deformations.
The main difference between minimal and extended decoupling schemes
is that the geometric deformation is applied only on the radial
metric component in the former case, leaving temporal coefficient as
an invariant quantity \cite{aa}. However, in the later scenario,
both temporal as well as radial potentials are deformed to covert
the field equations in a new reference frame \cite{aaa}. Moreover,
the MGD technique works as long as the interaction between the
matter sources is purely gravitational, implying that each fluid
source must be conserved individually. This is in contrast with the
extended decoupling approach in which the total fluid configuration
is conserved but the non-conservation phenomenon occurs when it
comes to individual systems of equations. The transfer of the energy
between different source is also allowed in this scenario. Further,
the deformation function plays a crucial role in the process of
gravitational decoupling. Its selection is based on the specific
characteristics of the problem and the intended simplification, with
the additional requirement of ensuring the spherical symmetry of the
solution. In the current setup, we have $\mathrm{t}_1\rightarrow
0,~\mathrm{t}_2\rightarrow \mathrm{T}$. Hence, Eq.\eqref{g16}
switches into
\begin{equation}\label{g17}
\nu_4\rightarrow\nu_1=\nu_4, \quad \nu_5\rightarrow
e^{-\nu_2}=\nu_5+\eta\mathrm{T},
\end{equation}
where $\mathrm{T}=\mathrm{T}(r)$. It must be kept in mind that such
a linear mapping does not bother the considered spherical symmetric.
We apply the transformation \eqref{g17} on the system
\eqref{g8}-\eqref{g10} to divide into two different sets
corresponding to $\eta=0$ and $1$. The first set corresponds to the
initial (isotropic) source and is given by
\begin{align}\label{g18}
&e^{-\nu_2}\left(\frac{\nu_2'}{r}-\frac{1}{r^2}\right)
+\frac{1}{r^2}=8\pi\mu+\nu_3\left(3\mu-P\right),\\\label{g19}
&e^{-\nu_2}\left(\frac{1}{r^2}+\frac{\nu_1'}{r}\right)
-\frac{1}{r^2}=8\pi P-\nu_3\left(\mu-3P\right),\\\label{g20}
&\frac{e^{-\nu_2}}{4}\left[\nu_1'^2-\nu_2'\nu_1'+2\nu_1''-\frac{2\nu_2'}{r}+\frac{2\nu_1'}{r}\right]
=8\pi P-\nu_3\left(\mu-3P\right).
\end{align}
The simultaneous solution of Eqs.\eqref{g18} and \eqref{g19} results
in the state variables as
\begin{align}\label{g18a}
\mu&=\frac{e^{-\nu_2}}{8r^2\big(\nu_3^2+6\pi\nu_3+8\pi^2\big)}\big[\nu_3
r \nu_1 '+(3 \nu_3 +8 \pi ) r \nu_2 '+2 (\nu_3 +4 \pi )
\big(e^{\nu_2 }-1\big)\big],\\\label{g19a}
P&=\frac{e^{-\nu_2}}{8r^2\big(\nu_3^2+6\pi\nu_3+8\pi^2\big)}\big[(3
\nu_3 +8 \pi ) r \nu_1 '+\nu_3  r \nu_2 '-2 (\nu_3 +4 \pi )
\big(e^{\nu_2 }-1\big)\big].
\end{align}
Contrariwise, the field equations representing the additional matter
distribution ($\mathfrak{D}^{\sigma}_{\omega}$) are obtained as
\begin{align}\label{g21}
&8\pi\mathfrak{D}_{0}^{0}=\frac{\mathrm{T}'}{r}+\frac{\mathrm{T}}{r^2},\\\label{g22}
&8\pi\mathfrak{D}_{1}^{1}=\mathrm{T}\left(\frac{\nu_1'}{r}+\frac{1}{r^2}\right),\\\label{g23}
&8\pi\mathfrak{D}_{2}^{2}=\frac{\mathrm{T}}{4}\left(2\nu_1''+\nu_1'^2+\frac{2\nu_1'}{r}\right)
+\mathrm{T}'\left(\frac{\nu_1'}{4}+\frac{1}{2r}\right).
\end{align}

It is interesting to note that both (seed and new) sources are
individually conserved, and hence, the energy's exchange between
them is not allowed in this scheme. Further, we successfully
decouple the set of equations \eqref{g8}-\eqref{g10} that makes
easier to solve both sets independently. There appear four unknowns
($\mu,P,\nu_1,\nu_2$) in Eqs.\eqref{g18a} and \eqref{g19a}, thus we
must require a well-behaved metric ansatz to equal the number of
unknowns to that of equations. Also, four unknowns
($\mathrm{T},\mathfrak{D}_{0}^{0},\mathfrak{D}_{1}^{1},\mathfrak{D}_{2}^{2}$)
are appeared in the system \eqref{g21}-\eqref{g23}, a unique
solution shall be obtained by adopting a constraint on
$\mathfrak{D}$-sector. Since the additional source makes the initial
matter source anisotropic, we identify the matter determinants as
follows
\begin{equation}\label{g13}
\tilde{\mu}=\mu-\eta\mathfrak{D}_{0}^{0},\quad
\tilde{P}_{r}=P+\eta\mathfrak{D}_{1}^{1}, \quad
\tilde{P}_{\bot}=P+\eta\mathfrak{D}_{2}^{2},
\end{equation}
which help to define the total anisotropic factor as
\begin{equation}\label{g14}
\tilde{\Pi}=\tilde{P}_{\bot}-\tilde{P}_{r}=\eta(\mathfrak{D}_{2}^{2}-\mathfrak{D}_{1}^{1}),
\end{equation}
verifying its disappearance for the case when $\eta=0$, i.e., the
effect of a new matter source is removed.

\section{Buchdahl Solution and Boundary Conditions}

This section is devoted to the formulation of a solution
corresponding to the first set given by \eqref{g18a} and
\eqref{g19a}. We have already discussed that a particular form of
metric potentials is required to solve the system. In this context,
Buchdahal ansatz \cite{42a} has attracted significant attention
among researchers. This assumption has proven valuable in
investigating almost all physically viable known super-dense star
models. Vaidya and Tikekar \cite{42} further refined the Buchdahl
ansatz and discussed spheroidal geometries for the 4-dimensional
hypersurface. Such a particular spheroidal condition has been proved
to be highly effective in obtaining exact solutions to the field
equations, a task that proves challenging in numerous other
scenarios. Kumar et al. \cite{58} extensively explored this
particular spacetime, focusing on charged compact objects coupled
with the isotropic fluid distribution. Additionally, Sharma et al.
\cite{63} determined the maximum attainable masses and radii for
various values of the surface density within the Vaidya-Tikekar
spacetime. Maurya et al. \cite{a} adopted this metric and discussed
eight different cases for Buchdahl's dimensionless parameter and
found all of them valid at every point within the interior geometry.
Various authors used this metric to produce viable as well stable
charged/uncharged interiors in the contexts of $\mathbb{GR}$ and
$f(\mathbb{R},\mathbb{T})$ theory \cite{b}-\cite{d}. This metric is
adopted as follows
\begin{eqnarray}\label{g33}
e^{\nu_1(r)}&=&\mathrm{C}_{1}\left[\big(1+\mathrm{C}_{3}r^2\big)^{\frac{3}{2}}+\mathrm{C}_{2}\big(5+2\mathrm{C}_{3}r^2\big)
\sqrt{2-\mathrm{C}_{3}r^2}\right]^2,\\\label{g34}
e^{\nu_2(r)}&=&\frac{2\big(1+\mathrm{C}_{3}r^2\big)}{2-\mathrm{C}_{3}r^2},\\\nonumber
\mu&=&\frac{3\mathrm{C}_{3}}{8(\nu_3+2\pi)(\nu_3+4\pi)\big(\mathrm{C}_{3}r^2+1\big)^2}
\big\{\big(\mathrm{C}_{3}r^2+1\big)\sqrt{2-\mathrm{C}_{3}^2r^4+\mathrm{C}_{3}r^2}\\\nonumber
&-&\mathrm{C}_{2}\big(\mathrm{C}_{3}r^2-2\big)\big(2\mathrm{C}_{3}r^2+5\big)\big\}^{-1}
\big[2 \big(\mathrm{C}_{3} r^2+1\big) \sqrt{2-\mathrm{C}_{3}^2 r^4+c
r^2} \\\nonumber &\times&\big\{3 \nu_3 +2 \pi \big(\mathrm{C}_{3}
r^2+3\big)\big\}-\mathrm{C}_{2} \big(\mathrm{C}_{3} r^2-2\big)
\big\{3 \nu_3 \big(4\mathrm{C}_{3} r^2+7\big)\\\label{g35} &+&4 \pi
\big(\mathrm{C}_{3} r^2+3\big) \big(2 \mathrm{C}_{3}
r^2+5\big)\big\}\big],\\\nonumber
P&=&\frac{3\mathrm{C}_{3}}{8(\nu_3+2\pi)(\nu_3+4\pi)\big(\mathrm{C}_{3}r^2+1\big)^2}
\big\{\mathrm{C}_{2}\big(\mathrm{C}_{3}r^2-2\big)\big(2\mathrm{C}_{3}r^2+5\big)\\\nonumber
&-&\big(\mathrm{C}_{3}r^2+1\big)\sqrt{2-\mathrm{C}_{3}^2r^4+\mathrm{C}_{3}r^2}\big\}^{-1}
\big[2 \big(\mathrm{C}_{3}r^2+1\big) \sqrt{2-\mathrm{C}_{3}^2
r^4+\mathrm{C}_{3}r^2}\\\nonumber
&\times&\big\{\nu_3\big(2\mathrm{C}_{3}r^2-3\big)+6\pi\big(\mathrm{C}_{3}r^2-1\big)\big\}-\mathrm{C}_{2}\big(\mathrm{C}_{3}
r^2-2\big) \big\{4 \big(\mathrm{C}_{3}r^2+1\big)\\\label{g36}
&\times&\big(2\nu_3\mathrm{C}_{3}r^2+\pi\big(6\mathrm{C}_{3}r^2+3\big)\big)-3
\nu_3 \big\}\big],
\end{eqnarray}
where $\mathrm{C}_{1},~\mathrm{C}_{2}$ and $\mathrm{C}_{3}$ are
unknown quantities that must be determined to perform the graphical
analysis.

In this perspective, the junction (or matching) conditions play a
vital role in the study of structural characteristics of a compact
object at some boundary surface ($\Sigma:~r=\mathrm{R}$). The
interior geometry now becomes in terms of ansatz \eqref{g33} and
\eqref{g34} as
\begin{align}\nonumber
ds_-^2&=-\mathrm{C}_{1}\left[\big(1+\mathrm{C}_{3}r^2\big)^{\frac{3}{2}}+\mathrm{C}_{2}\big(5+2\mathrm{C}_{3}r^2\big)
\sqrt{2-\mathrm{C}_{3}r^2}\right]^2dt^2\\\label{g36aa}
&+\frac{2\big(1+\mathrm{C}_{3}r^2\big)}{2-\mathrm{C}_{3}r^2}
dr^2+r^2d\theta^2+r^2\sin^2\theta d\vartheta^2,
\end{align}
whose corresponding exterior spacetime is described by the
Schwarzschild metric. This is given as follows
\begin{equation}\label{g25}
ds_+^2=-\frac{r-2\mathrm{M}}{r}dt^2+\frac{r}{r-2\mathrm{M}}dr^2+
r^2d\theta^2+r^2\sin^2\theta d\vartheta^2,
\end{equation}
where $\mathrm{M}$ shows the total mass. The first fundamental form
of these constraints provide continuity of the metric potentials at
the hypersurface as
\begin{equation}\nonumber
g_{tt}^-~{_=^\Sigma}~g_{tt}^+, \quad g_{rr}^-~{_=^\Sigma}~g_{rr}^+,
\end{equation}
providing the following expressions by employing the metrics
\eqref{g36aa} and \eqref{g25} as
\begin{eqnarray}\label{g36ab}
\frac{\mathrm{R}-2\mathrm{M}}{\mathrm{R}}&=&\mathrm{C}_{1}\left[\big(1+\mathrm{C}_{3}\mathrm{R}^2\big)^{\frac{3}{2}}
+\mathrm{C}_{2}\big(5+2\mathrm{C}_{3}\mathrm{R}^2\big)\sqrt{2-\mathrm{C}_{3}\mathrm{R}^2}\right]^2,\\\label{g36ac}
\frac{\mathrm{R}}{\mathrm{R}-2\mathrm{M}}&=&\frac{2\big(1+\mathrm{C}_{3}\mathrm{R}^2\big)}{2-\mathrm{C}_{3}\mathrm{R}^2}.
\end{eqnarray}
\begin{table}
\scriptsize \centering \caption{Unknown triplet
($\mathrm{C}_{1},\mathrm{C}_{2},\mathrm{C}_{3}$) for different
values of the parameter $\nu_3$ corresponding to a star LMC X-4.}
\label{Table1} \vspace{+0.07in} \setlength{\tabcolsep}{0.95em}
\begin{tabular}{ccccccc}
\hline\hline $\nu_3$ & 0 & 0.25 & 0.5 & 0.75 & 1 & 1.25
\\\hline $\mathrm{C}_{1}\times 10^{-2}$ & 3.6039 & 3.4065 & 3.2253 & 3.0585 &
2.9047 & 2.7626
\\\hline $\mathrm{C}_{2}\times 10^{-1}$ & 3.6616 & 3.8256 & 3.9892 & 4.1524 &
4.3153 & 4.4777
\\\hline
$\mathrm{C}_{3}\times 10^{-3}~(km^{-2})$ & 4.5616 & 4.5616 & 4.5616 & 4.5616 & 4.5616 & 4.5616\\
\hline\hline
\end{tabular}
\end{table}

Since we have two equations and three unknowns, we need one more
constraint. For this, we use the second fundamental form that claims
that the isotropic pressure disappears at the boundary, i.e.,
$P~{_=^\Sigma}~0$. Thus, Eq.\eqref{g36} provides
\begin{align}\nonumber
&2 \big(\mathrm{C}_{3}r^2+1\big) \sqrt{2-\mathrm{C}_{3}^2
r^4+\mathrm{C}_{3}r^2}\big\{\nu_3\big(2\mathrm{C}_{3}r^2-3\big)+6\pi\big(\mathrm{C}_{3}r^2-1\big)\big\}\\\label{g36ad}
&-\mathrm{C}_{2}\big(\mathrm{C}_{3} r^2-2\big) \big\{4
\big(\mathrm{C}_{3}r^2+1\big)\big(2\nu_3\mathrm{C}_{3}r^2+\pi\big(6\mathrm{C}_{3}r^2+3\big)\big)-3
\nu_3 \big\}=0.
\end{align}
By solving Eqs.\eqref{g36ab}-\eqref{g36ad} simultaneously, we get
the triplet $(\mathrm{C}_{1},\mathrm{C}_{2},\mathrm{C}_{3})$ as
\begin{eqnarray}\label{g37}
\mathrm{C}_{1}&=&\frac{1}{3\mathrm{F}^2}\bigg(\frac{\mathrm{R}-2\mathrm{M}}{\mathrm{R}}\bigg),\\\label{g38}
\mathrm{C}_{2}&=&\frac{\mathrm{R}(3\mathrm{R}-4\mathrm{M})\sqrt{\frac{2\mathrm{R}(\mathrm{R}-2\mathrm{M})}
{(4\mathrm{M}-3\mathrm{R})^2}}\big\{4\mathrm{M}(5\nu_3+12\pi)-9\mathrm{R}(\nu_3+2\pi)\big\}}
{(2\mathrm{M}-\mathrm{R})\big\{8\mathrm{MR}(7\nu_3+6\pi)
-16\nu_3\mathrm{M}^2-9\mathrm{R}^2(\nu_3-4\pi)\big\}},\\\label{g38a}
\mathrm{C}_{3}&=&-\frac{4\mathrm{M}}{\mathrm{R}^2(4\mathrm{M}-3\mathrm{R})},
\end{eqnarray}
where $\mathrm{F}$ is provided by
\begin{align}\label{g38b}
\mathrm{F}&=3\bigg(\frac{\mathrm{R}}{3\mathrm{R}-4\mathrm{M}}\bigg)^{\frac{3}{2}}-\frac{6R\big\{4\mathrm{M}(5\nu_3+12\pi)
-9\mathrm{R}(\nu_3+2\pi)\big\}}{8\mathrm{MR}(7\nu_3+6\pi)-16\nu_3\mathrm{M}^2-9\mathrm{R}^2(\nu_3-4\pi)}\\
&\times\bigg[\frac{(4\mathrm{M}-5\mathrm{R})}{(2\mathrm{M}-\mathrm{R})}\sqrt{\frac{2\mathrm{M}-\mathrm{R}}
{4\mathrm{M}-3\mathrm{R}}}\sqrt{\frac{\mathrm{R}(\mathrm{R}-2\mathrm{M})}{(4\mathrm{M}-3\mathrm{R})^2}}\bigg].
\end{align}
We use the calculated values of the mass and radius as
$\mathrm{M}=1.04 \pm 0.09 M_{\bigodot}$ and $\mathrm{R}=8.301 \pm
0.2 km$, respectively, of a compact star candidate LMC X-4
\cite{42aa} to evaluate the above constant triplet. Table \textbf{1}
provides these values corresponding to different choices of the
model parameter. It is observed that the constant $\mathrm{C}_{1}$
($\mathrm{C}_{2}$) is in inverse (direct) relation with the
parameter $\nu_3$, however, the unknown $\mathrm{C}_{3}$ is not
dependent on $\nu_3$.

\section{Physical Acceptability Conditions of a Compact Model}

There are several criteria in the literature to check physical
acceptability of compact structures \cite{ab}-\cite{af}. Some other
works are \cite{aza}-\cite{azd}. They are highlighted as follows.
\begin{itemize}
\item One of the fundamental aspects is the existence of geometric singularities inside the
star. The behavior of the metric components should be positively
increasing and free from singularity in an acceptable
self-gravitating interior. The metric coefficients \eqref{g33} and
\eqref{g34} in the core of star become
\begin{align}\nonumber
e^{\nu_1(r)}|_{r=0}=\mathrm{C}_{1}\left[1+5\sqrt{2}\mathrm{C}_{2}\right]^2,\quad
e^{\nu_2(r)}|_{r=0}=1,
\end{align}
and their first derivatives are
\begin{align}\nonumber
(e^{\nu_1(r)})'&=\frac{6\mathrm{C}_{1}\mathrm{C}_{3}r\mathrm{f}(r)\big\{\big(1+\mathrm{C}_{3}r^2\big)^{\frac{3}{2}}+\mathrm{C}_{2}\big(5+2\mathrm{C}_{3}r^2\big)
\sqrt{2-\mathrm{C}_{3}r^2}\big\}}{\sqrt{2-\mathrm{C}_{3}r^2}},\\\nonumber
(e^{\nu_2(r)})'&=\frac{12\mathrm{C}_{3}r}{\big(2-\mathrm{C}_{3}r^2\big)^2},
\end{align}
where
$\mathrm{f}(r)=\mathrm{C}_{2}-2\mathrm{C}_{2}\mathrm{C}_{3}r^2+\sqrt{2+\mathrm{C}_{3}r^2-\mathrm{C}_{3}^2r^4}$.
We notice that the above both derivatives disappear at the center
implying the regularity of these potentials. Figure \textbf{1}
reveals that they are minimum at $r=0$ and increasing outwards to
reach their maximum at the spherical surface.

\item The profile of matter determinants such as energy density and
pressure components should be finite and positive everywhere.
Further, they must reach their maximum (minimum) at $r=0$
($\Sigma:~r=\mathrm{R}$). Likewise, the decreasing trend of these
variables towards the boundary can also be assured if their first
derivative disappears at $r=0$ and negative outwards.

\item The mass of the spherical stellar structure is defined by
\begin{equation}\label{g39}
\mathrm{m}(r)=\frac{1}{2}\int_{0}^{\mathrm{R}}\mathrm{w}^2\mu
d\mathrm{w}.
\end{equation}
The mass-radius ratio, known as the compactness, measures that how
tightly the particles are bound with each other in a
self-gravitating system. Mathematically, we have
\begin{equation}\label{g40}
\zeta(r)=\frac{\mathrm{m}(r)}{r}.
\end{equation}
Its value must be less than $\frac{4}{9}$ everywhere in a spherical
interior \cite{42a}. We can also describe the surface redshift in
terms of mass or compactness of a star. This is given as
\begin{equation}\label{g41}
\mathrm{z}(r)=\frac{1-\sqrt{1-2\zeta(r)}}{\sqrt{1-2\zeta(r)}}.
\end{equation}
Since we are discussing anisotropic matter distribution, the
redshift reaches its maximum by $5.211$ at the boundary surface to
get a feasible model \cite{42b}.

\item Another key feature to check feasibility of the stellar model
is the energy conditions which are in fact linear combinations of
the matter determinants. These are given as follows
\begin{eqnarray}\nonumber
&&\mu \geq 0, \quad \mu+P_{\bot} \geq 0,\\\nonumber &&\mu+P_{r} \geq
0, \quad \mu-P_{\bot} \geq 0,\\\label{g50} &&\mu-P_{r} \geq 0, \quad
\mu+2P_{\bot}+P_{r} \geq 0.
\end{eqnarray}
Among these conditions, the dominant energy bounds (i.e.,
$\mu-P_{\bot} \geq 0$ and $\mu-P_{r} \geq 0$) are the most important
because they demand $\mu \geq P_{\bot}$ and $\mu \geq P_{r}$
everywhere.

\item Different techniques have been suggested in the
literature to check the stability of the compact stars, one of them
is based on the sound speed whose radial and tangential components
are $v_{sr}^{2}=\frac{dP_{r}}{d\mu}$ and
$v_{s\bot}^{2}=\frac{dP_{\bot}}{d\mu}$, respectively. Abreu et al.
\cite{42bb} suggested that if the speed of light is greater than
that of sound (i.e., $0 < v_{sr}^{2},~ v_{s\bot}^{2} < 1$), the
causality would be preserved in a system. Likewise, Herrera
\cite{42ba} proposed that if the total radial force changes its sign
throughout the evolution, then the idea of cracking would occur in
the interior fluid. It must be avoided to obtain a stable model. He
provided that the cracking cannot be occurred only if $-1 <
v_{s\bot}^{2}-v_{sr}^{2} < 0$ holds.
\end{itemize}
\begin{figure}\center
\epsfig{file=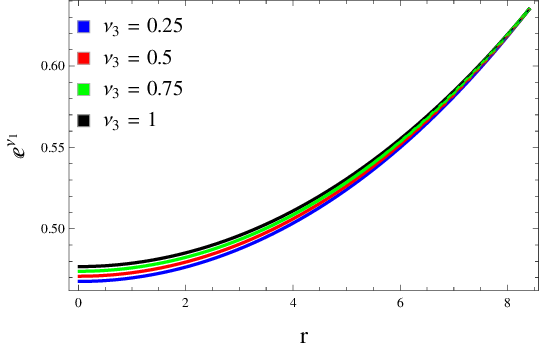,width=0.4\linewidth}\epsfig{file=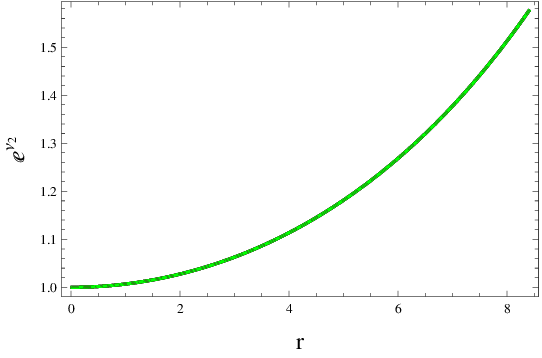,width=0.4\linewidth}
\caption{Buchdahl metric ansatz \eqref{g33} and \eqref{g34} for
$\eta=0.1$ (solid) and $0.3$ (dotted).}
\end{figure}

\section{Formation of Different Anisotropic Models}

Here we consider two sources characterizing different matter
distributions, the unknowns are increased in the field equations.
Therefore, we choose three different constraints to make such
differential equations solvable. They are given by
\begin{itemize}
\item Model I: Density-like constraint
\item Model II: Pressure-like constraint
\item Model III: A linear equation of state
\end{itemize}

\subsection{Model I}

We consider a constraint depending on the energy density of the
original and anisotropic source to obtain solution to the total
interior fluid. This is given as follows \cite{42baba}
\begin{equation}\label{g51}
\mu=\mathfrak{D}_{0}^{0}.
\end{equation}
We use Eqs.\eqref{g18a} and \eqref{g22} in \eqref{g51}, and obtain
the differential equation as
\begin{align}\nonumber
&\frac{1}{8\pi}\bigg\{\frac{\mathrm{T}'(r)}{r}+\frac{\mathrm{T}(r)}{r^2}\bigg\}
-\frac{e^{-\nu_2}}{8r^2\big(\nu_3^2+6\pi\nu_3+8\pi^2\big)}\\\label{g52}
&\times\big[\nu_3 r \nu_1 '+(3 \nu_3 +8 \pi ) r \nu_2 '+2 (\nu_3 +4
\pi ) \big(e^{\nu_2 }-1\big)\big]=0.
\end{align}
In terms of Buchdahl's ansatz \eqref{g33} and \eqref{g34}, the above
equation takes the form
\begin{align}\nonumber
&\frac{1}{8\pi}\bigg\{\frac{\mathrm{T}'(r)}{r}+\frac{\mathrm{T}(r)}{r^2}\bigg\}-3
\mathrm{C}_3\big[8 \big(\nu _3+2 \pi \big) \big(\nu _3+4 \pi \big)
\big(\mathrm{C}_3 r^2+1\big)^2 \big\{\big(\mathrm{C}_3
r^2+1\big)\\\nonumber &\times\sqrt{\mathrm{C}_3 r^2+2-\mathrm{C}_3^2
r^4}-\mathrm{C}_2 \big(\mathrm{C}_3 r^2-2\big) \big(2 \mathrm{C}_3
r^2+5\big)\big\}\big]^{-1}\big[2 \big(\mathrm{C}_3
r^2+1\big)\\\nonumber & \sqrt{\mathrm{C}_3 r^2+2-\mathrm{C}_3^2 r^4}
\big(2 \pi \big(\mathrm{C}_3 r^2+3\big)+3 \nu _3\big)-\mathrm{C}_2
\big(\mathrm{C}_3 r^2-2\big) \big\{3 \nu _3\big(4 \mathrm{C}_3
r^2+7\big)\\\label{g53}& +4 \pi \big(\mathrm{C}_3 r^2+3\big) \big(2
\mathrm{C}_3 r^2+5\big)\big\}\big]=0.
\end{align}
The above equation contains one unknown \big(i.e., the deformation
function $\mathrm{T}(r)$\big). However, the exact solution of this
equation is not possible because of the terms appearing in the
square root. Therefore, we use the numerical integration to make
this function known. For this purpose, we use the initial condition
$\mathrm{T}(0)=0$ by providing the range of the radius of the
considered star. The corresponding deformed $g_{rr}^{-1}$ component
can be calculated by
\begin{equation}\label{g54}
e^{-\nu_2(r)}=\frac{2-\mathrm{C}_{3}r^2+2\eta\mathrm{T}\big(1+\mathrm{C}_{3}r^2\big)}{2\big(1+\mathrm{C}_{3}r^2\big)}.
\end{equation}

We perform the graphical analysis of the obtained deformation
function and its corresponding matter determinants, anisotropy,
energy conditions and some other parameters for the considered
modified model \eqref{g7a}. Moreover, we choose $\eta=0.1,~0.3$ and
$\nu_3=0.25,~0.5,~0.75,~1$ to study the effect of decoupling
strategy and the modified gravity on the interior of a compact star.
The deformation function obtained from Eq.\eqref{g53} does not
depend on the decoupling parameter, thus its variation with respect
to different values of the model parameter is plotted in Figure
\textbf{2}. It is observed that this function possesses increasing
trend outwards, however, it takes smaller values for increasing
$\nu_3$.

The corresponding matter variables such as effective energy density,
effective radial/tangential pressures and anisotropy are calculated
through Eqs.\eqref{g13} and \eqref{g14}. The trend of these
variables is plotted in Figure \textbf{3}, indicating an acceptable
behavior. We notice that the energy density (pressure components)
decreases (increase) with the increment in $\eta$. However, they
decrease by increasing the model parameter. Such a behavior can be
observed by the numerical values provided in Tables \textbf{2} and
\textbf{3}. Further, the zero radial pressure at the spherical
boundary is observed for each parametric value (upper right plot).
The anisotropic factor, in this case, becomes null in the core and
increases, otherwise. The parameters in Figure \textbf{4} possess
increasing and acceptable profile throughout. The energy density in
Eq.\eqref{g39} corresponds to the total matter source involving
modified corrections, thus the mass function is dependent on the
decoupling and model parameters. The smaller values of both
parameters $\nu_3$ and $\eta$ provide more massive interiors (upper
left plot). Figure \textbf{5} shows the dominant energy bounds,
displaying positive profile and thus leading to a viable solution.
Figure \textbf{6} (lower plot) declares that our resulting model I
is physically unstable because there occurs a cracking in the
interior for every parametric choice.
\begin{table}
\scriptsize \centering \caption{Values of central density, surface
density, central pressure, surface compactness and surface redshift
corresponding to $\eta=0.1$ for Model I.} \label{Table2}
\vspace{+0.07in} \setlength{\tabcolsep}{0.95em}
\begin{tabular}{ccccccc}
\hline\hline $\nu_3$ & 0 & 0.25 & 0.5 & 0.75
\\\hline $\mu_c~(gm/cm^3)$ & 1.0625$\times$10$^{15}$ & 1.0316$\times$10$^{15}$ & 1.0058$\times$10$^{15}$ &
9.7997$\times$10$^{14}$
\\\hline $\mu_s~(gm/cm^3)$ & 6.6785$\times$10$^{14}$ & 6.5501$\times$10$^{14}$ & 6.3695$\times$10$^{14}$ &
6.1621$\times$10$^{14}$
\\\hline $P_c ~(dyne/cm^2)$ & 1.2758$\times$10$^{35}$ & 1.2373$\times$10$^{35}$ & 1.1938$\times$10$^{35}$ &
1.1561$\times$10$^{35}$
\\\hline $\zeta_s$ & 0.178 & 0.174 & 0.168 & 0.163
\\\hline
$z_s$ & 0.245 & 0.237 & 0.231 & 0.221\\
\hline\hline
\end{tabular}
\end{table}
\begin{table}
\scriptsize \centering \caption{Values of central density, surface
density, central pressure, surface compactness and surface redshift
corresponding to $\eta=0.3$ for Model I.} \label{Table3}
\vspace{+0.07in} \setlength{\tabcolsep}{0.95em}
\begin{tabular}{ccccccc}
\hline\hline $\nu_3$ & 0 & 0.25 & 0.5 & 0.75
\\\hline $\mu_c~(gm/cm^3)$ & 1.0548$\times$10$^{15}$ & 1.0264$\times$10$^{15}$ & 9.9816$\times$10$^{14}$ &
9.7221$\times$10$^{14}$
\\\hline $\mu_s~(gm/cm^3)$ & 6.6263$\times$10$^{14}$ & 6.4979$\times$10$^{14}$ & 6.3427$\times$10$^{14}$ &
6.1367$\times$10$^{14}$
\\\hline $P_c ~(dyne/cm^2)$ & 1.2878$\times$10$^{35}$ & 1.2481$\times$10$^{35}$ & 1.2001$\times$10$^{35}$ &
1.1689$\times$10$^{35}$
\\\hline $\zeta_s$ & 0.142 & 0.137 & 0.133 & 0.129
\\\hline
$z_s$ & 0.179 & 0.174 & 0.168 & 0.164\\
\hline\hline
\end{tabular}
\end{table}
\begin{figure}\center
\epsfig{file=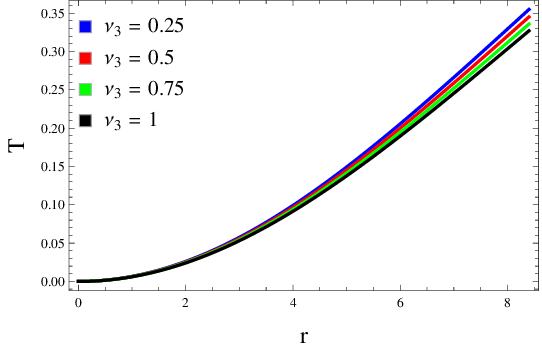,width=0.4\linewidth}
\caption{Deformation function corresponding to model I.}
\end{figure}

\subsection{Model II}
\begin{figure}\center
\epsfig{file=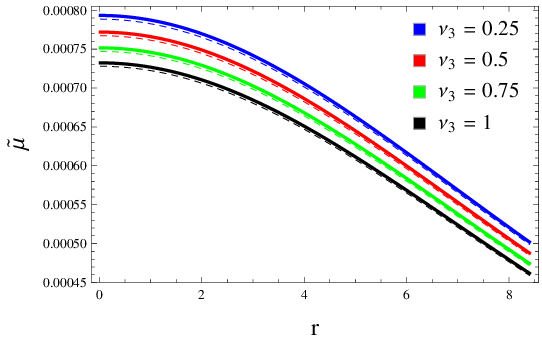,width=0.4\linewidth}\epsfig{file=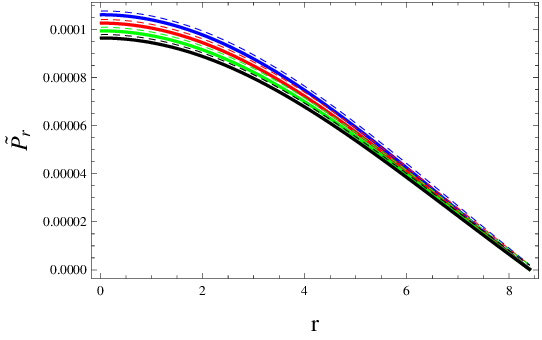,width=0.4\linewidth}
\epsfig{file=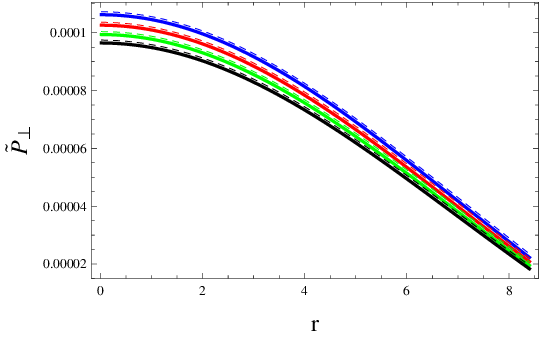,width=0.4\linewidth}\epsfig{file=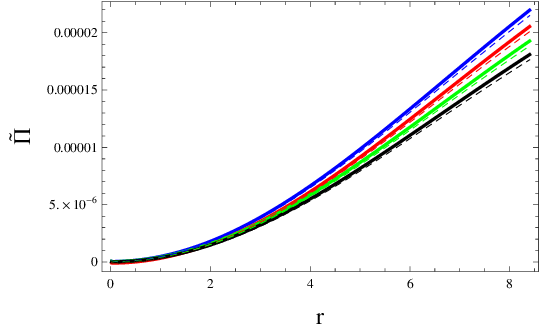,width=0.4\linewidth}
\caption{Physical variables and anisotropy for $\eta=0.1$ (solid)
and $0.3$ (dotted) corresponding to model I.}
\end{figure}
\begin{figure}\center
\epsfig{file=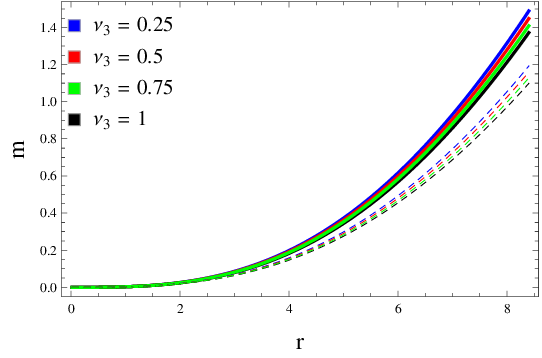,width=0.4\linewidth}\epsfig{file=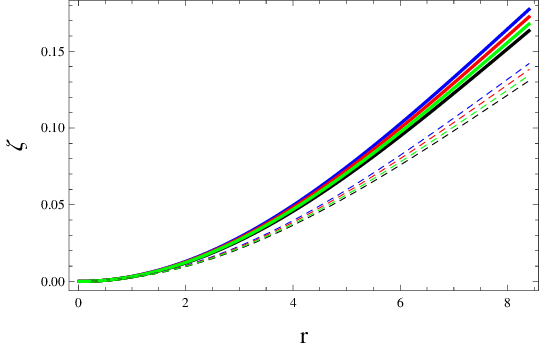,width=0.4\linewidth}
\epsfig{file=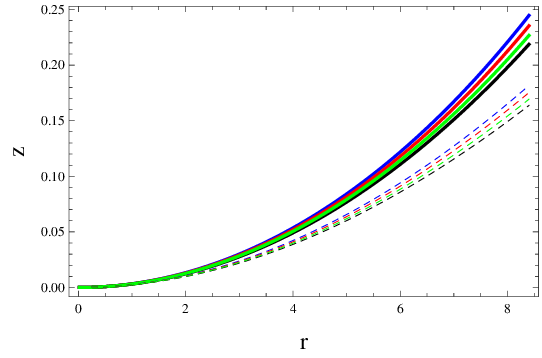,width=0.4\linewidth} \caption{Different
parameters for $\eta=0.1$ (solid) and $0.3$ (dotted) corresponding
to model I.}
\end{figure}
\begin{figure}\center
\epsfig{file=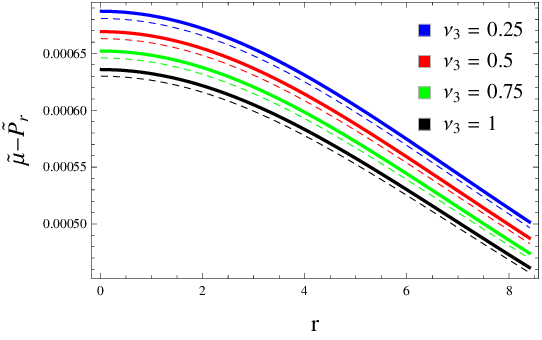,width=0.4\linewidth}\epsfig{file=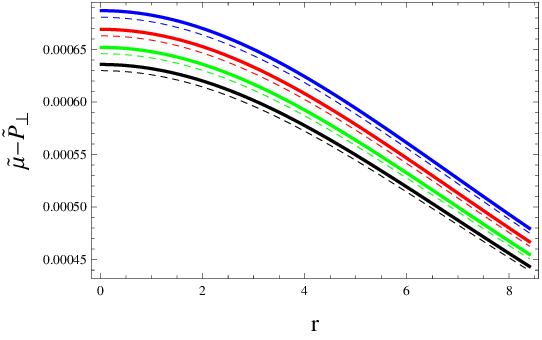,width=0.4\linewidth}
\caption{Dominant energy bounds for $\eta=0.1$ (solid) and $0.3$
(dotted) corresponding to model I.}
\end{figure}
\begin{figure}\center
\epsfig{file=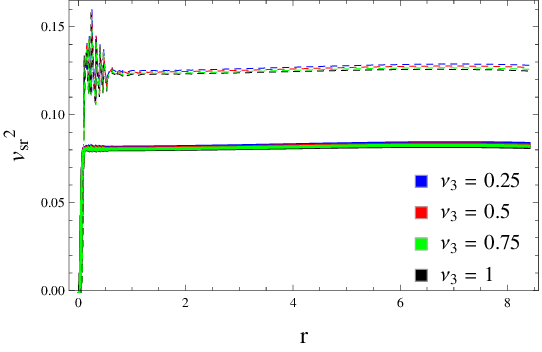,width=0.4\linewidth}\epsfig{file=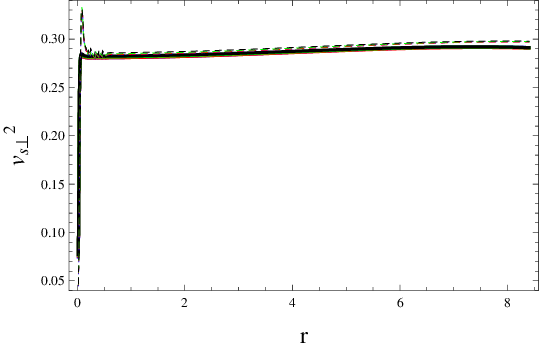,width=0.4\linewidth}
\epsfig{file=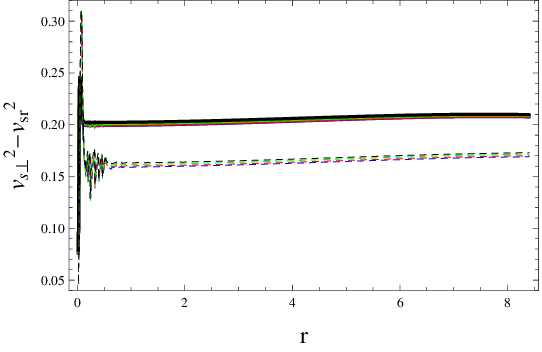,width=0.4\linewidth} \caption{Stability
criteria for $\eta=0.1$ (solid) and $0.3$ (dotted) corresponding to
model I.}
\end{figure}

In this subsection, we construct our second model by choosing the
following constraint \cite{42babb}
\begin{equation}\label{g55}
P=\mathfrak{D}_{1}^{1}.
\end{equation}
Using Eqs.\eqref{g19a} and \eqref{g23} in the above equation, we
have
\begin{align}\nonumber
&\frac{\mathrm{T}(r)}{8\pi}\bigg\{\frac{\nu_1'}{r}+\frac{1}{r^2}\bigg\}
-\frac{e^{-\nu_2}}{8r^2\big(\nu_3^2+6\pi\nu_3+8\pi^2\big)}\\\label{g56}
&\times\big[(3 \nu_3 +8 \pi ) r \nu_1 '+\nu_3  r \nu_2 '-2 (\nu_3 +4
\pi ) \big(e^{\nu_2 }-1\big)\big]=0,
\end{align}
that, in terms of the ansatz \eqref{g33} and \eqref{g34}, leading to
\begin{align}\nonumber
&\frac{\mathrm{T}(r)}{8\pi}\bigg\{\frac{6 \mathrm{C}_3
\big(\mathrm{C}_2-2 \mathrm{C}_2 \mathrm{C}_3
r^2+\sqrt{\mathrm{C}_3r^2+2-\mathrm{C}_3^2
r^4}\big)}{\big(\mathrm{C}_3 r^2+1\big) \sqrt{\mathrm{C}_3
r^2+2-\mathrm{C}_3^2 r^4}-\mathrm{C}_2 \big(\mathrm{C}_3 r^2-2\big)
\big(2\mathrm{C}_3 r^2+5\big)}+\frac{1}{r^2}\bigg\}\\\nonumber&-3
\mathrm{C}_3\big[8 \big(\nu _3+2 \pi \big) \big(\nu _3+4 \pi \big)
\big(\mathrm{C}_3 r^2+1\big)^2 \big\{\big(\mathrm{C}_3
r^2+1\big)\sqrt{\mathrm{C}_3 r^2+2-\mathrm{C}_3^2 r^4}\\\nonumber
&-\mathrm{C}_2 \big(\mathrm{C}_3 r^2-2\big) \big(2 \mathrm{C}_3
r^2+5\big)\big\}\big]^{-1}\big[2 \big(\mathrm{C}_3
r^2+1\big)\sqrt{\mathrm{C}_3 r^2+2-\mathrm{C}_3^2 r^4} \big\{\nu _3
\\\nonumber &\times \big(2 \mathrm{C}_3 r^2-3\big)+6 \pi
\big(\mathrm{C}_3 r^2-1\big)\big\}+\mathrm{C}_2 \big(\mathrm{C}_3
r^2-2\big) \big\{3 \nu _3-4 \big(\mathrm{C}_3
r^2+1\big)\\\label{g57} &\times \big(2 \mathrm{C}_3 \nu _3 r^2+\pi
\big(6 \mathrm{C}_3 r^2+3\big)\big)\big\}\big]=0.
\end{align}
The above equation provides the deformation function as
\begin{align}\nonumber
&\mathrm{T}(r)=\frac{3\pi\mathrm{C}_3r^2}{\big(\nu_3+2\pi\big)\big(\nu_3+4\pi\big)\big(\mathrm{C}_3r^2+1\big)^2}
\big[\mathrm{C}_2 \big\{\mathrm{C}_3 r^2 \big(14 \mathrm{C}_3
r^2-5\big)-10\big\}\\\nonumber &+\sqrt{\mathrm{C}_3
r^2+2-\mathrm{C}_3^2 r^4} \big(-7 \mathrm{C}_3
r^2-1\big)\big]^{-1}\big[2 \big(\mathrm{C}_3 r^2+1\big)
\sqrt{\mathrm{C}_3 r^2+2-\mathrm{C}_3^2 r^4}\\\nonumber &\times
\big\{\nu _3 \big(2 \mathrm{C}_3 r^2-3\big)+6 \pi \big(\mathrm{C}_3
r^2-1\big)\big\}-\mathrm{C}_2 \big(\mathrm{C}_3 r^2-2\big) \big\{4
\big(\mathrm{C}_3 r^2+1\big)\\\label{g58} &\times \big(2
\mathrm{C}_3 \nu _3 r^2+\pi \big(6 \mathrm{C}_3 r^2+3\big)\big)-3
\nu _3\big\}\big].
\end{align}

Figure \textbf{7} exhibits the plot of the above radial deformation
function for the considered pressure-like constraint with respect to
the chosen values of parameters. We observe that this function
initially increases for all values of model parameter, i.e., $\nu_3
\in (0,1)$, and then decreases to reach its minimum at the
hypersurface. The effective matter determinants along with
anisotropic factor corresponding to the above deformation function
can be obtained by making use of Eqs.\eqref{g13} and \eqref{g14}. We
also examine the profile of such effective quantities in Figure
\textbf{8}. We find from the upper left plot that the energy density
possesses the same (opposite) behavior as that of the first model
near the center (spherical boundary). Tables \textbf{4} and
\textbf{5} provide the numerical values that confirm such profile of
these matter determinants. Further, both pressure ingredients and
anisotropy show a consistent behavior. The radial pressure and
anisotropic factor are found to be null at the boundary and center,
respectively (right plots).
\begin{table}
\scriptsize \centering \caption{Values of central density, surface
density, central pressure, surface compactness and surface redshift
corresponding to $\eta=0.1$ for Model II.} \label{Table4}
\vspace{+0.07in} \setlength{\tabcolsep}{0.95em}
\begin{tabular}{ccccccc}
\hline\hline $\nu_3$ & 0 & 0.25 & 0.5 & 0.75
\\\hline $\mu_c~(gm/cm^3)$ & 1.0193$\times$10$^{15}$ & 9.9187$\times$10$^{14}$ & 9.6659$\times$10$^{14}$ &
9.4371$\times$10$^{14}$
\\\hline $\mu_s~(gm/cm^3)$ & 6.8283$\times$10$^{14}$ & 6.6451$\times$10$^{14}$ & 6.4391$\times$10$^{14}$ &
6.3026$\times$10$^{14}$
\\\hline $P_c ~(dyne/cm^2)$ & 1.3924$\times$10$^{35}$ & 1.3527$\times$10$^{35}$ & 1.3142$\times$10$^{35}$ &
1.2649$\times$10$^{35}$
\\\hline $\zeta_s$ & 0.179 & 0.175 & 0.169 & 0.162
\\\hline
$z_s$ & 0.245 & 0.235 & 0.229 & 0.219\\
\hline\hline
\end{tabular}
\end{table}
\begin{table}
\scriptsize \centering \caption{Values of central density, surface
density, central pressure, surface compactness and surface redshift
corresponding to $\eta=0.3$ for Model II.} \label{Table5}
\vspace{+0.07in} \setlength{\tabcolsep}{0.95em}
\begin{tabular}{ccccccc}
\hline\hline $\nu_3$ & 0 & 0.25 & 0.5 & 0.75
\\\hline $\mu_c~(gm/cm^3)$ & 9.3462$\times$10$^{14}$ & 9.0946$\times$10$^{14}$ & 8.8659$\times$10$^{14}$ &
8.6598$\times$10$^{14}$
\\\hline $\mu_s~(gm/cm^3)$ & 7.1026$\times$10$^{14}$ & 6.9193$\times$10$^{14}$ & 6.7133$\times$10$^{14}$ &
6.5541$\times$10$^{14}$
\\\hline $P_c ~(dyne/cm^2)$ & 1.6545$\times$10$^{35}$ & 1.5956$\times$10$^{35}$ & 1.5571$\times$10$^{35}$ &
1.4982$\times$10$^{35}$
\\\hline $\zeta_s$ & 0.169 & 0.168 & 0.163 & 0.161
\\\hline
$z_s$ & 0.238 & 0.227 & 0.218 & 0.215\\
\hline\hline
\end{tabular}
\end{table}

Figure \textbf{9} displays the variation in multiple factors with
respect to $r$ and parametric values, and we find them consistent
with observed data. Since we observe the matter triplet
$(\tilde{\mu},\tilde{P}_r,\tilde{P_\bot})$ to be positive
everywhere, the dominant energy bounds are only needed to check. We
plot them in Figure \textbf{10} and deduce that the corresponding
solution is viable. We also check the stability of the resulting
solution in Figure \textbf{11}, showing that our model II becomes
stable through both criteria (sound speed and cracking approach).
\begin{figure}\center
\epsfig{file=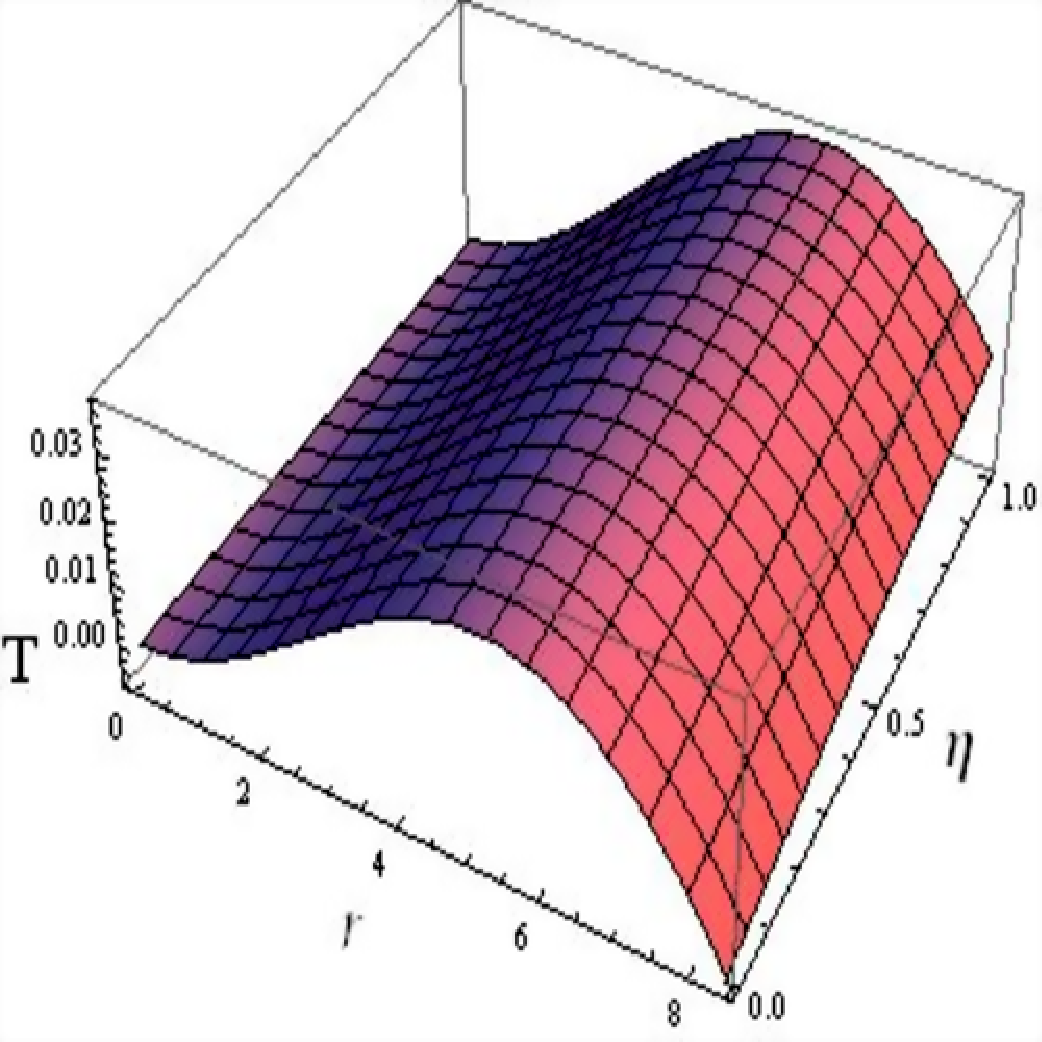,width=0.4\linewidth}
\caption{Deformation function corresponding to model II.}
\end{figure}
\begin{figure}\center
\epsfig{file=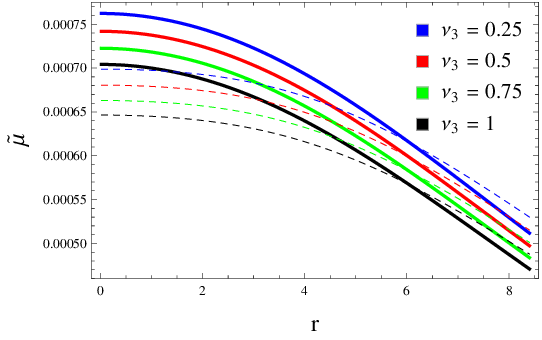,width=0.4\linewidth}\epsfig{file=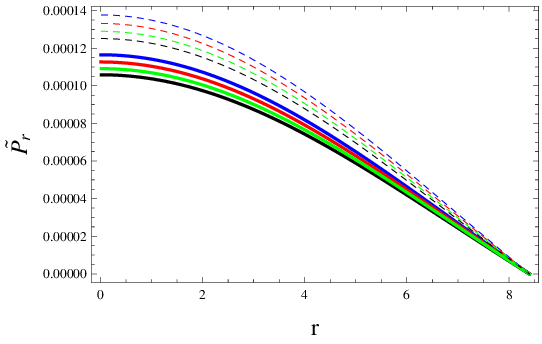,width=0.4\linewidth}
\epsfig{file=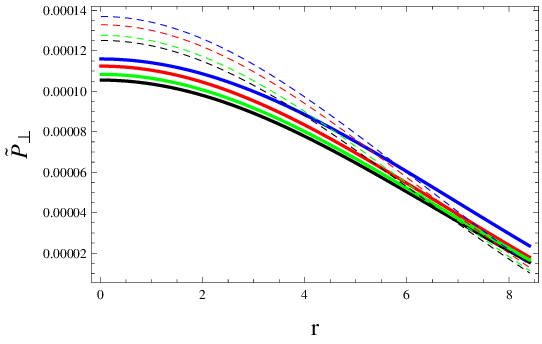,width=0.4\linewidth}\epsfig{file=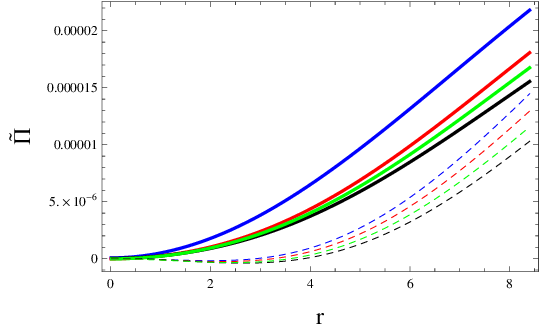,width=0.4\linewidth}
\caption{Physical variables and anisotropy for $\eta=0.1$ (solid)
and $0.3$ (dotted) corresponding to model II.}
\end{figure}
\begin{figure}\center
\epsfig{file=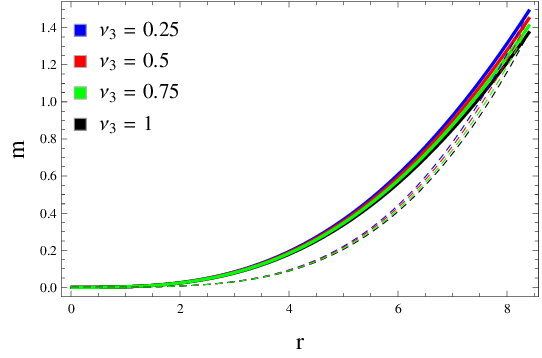,width=0.4\linewidth}\epsfig{file=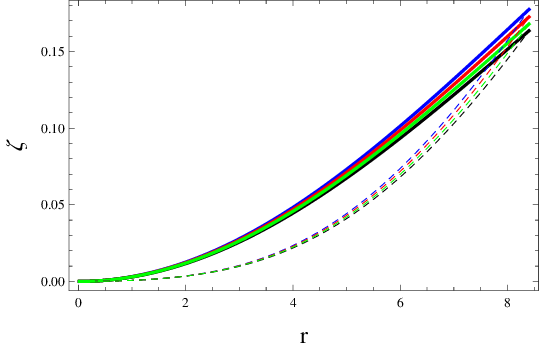,width=0.4\linewidth}
\epsfig{file=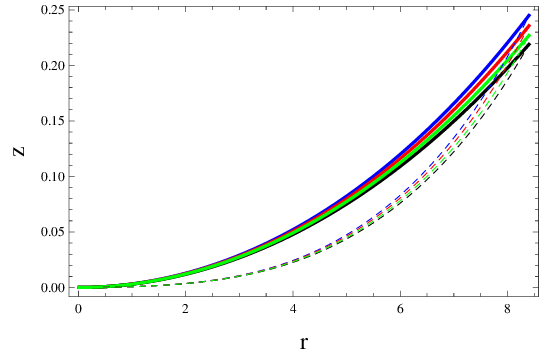,width=0.4\linewidth} \caption{Different
parameters for $\eta=0.1$ (solid) and $0.3$ (dotted) corresponding
to model II.}
\end{figure}
\begin{figure}\center
\epsfig{file=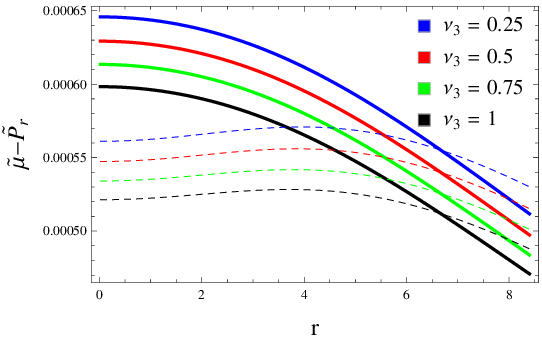,width=0.4\linewidth}\epsfig{file=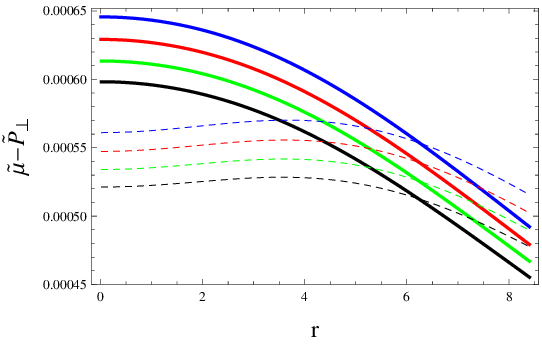,width=0.4\linewidth}
\caption{Dominant energy bounds for $\eta=0.1$ (solid) and $0.3$
(dotted) corresponding to model II.}
\end{figure}
\begin{figure}\center
\epsfig{file=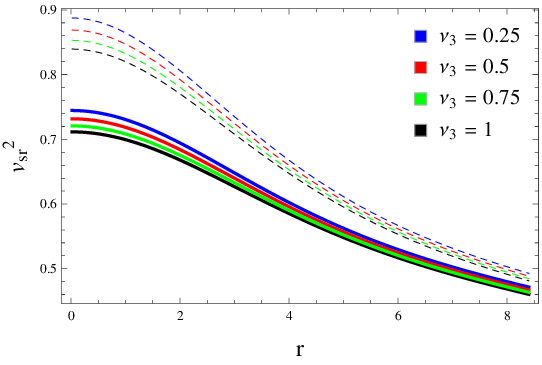,width=0.4\linewidth}\epsfig{file=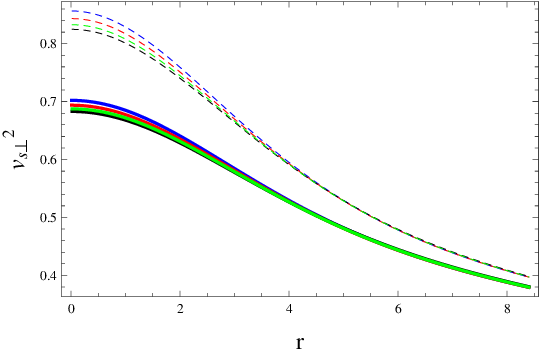,width=0.4\linewidth}
\epsfig{file=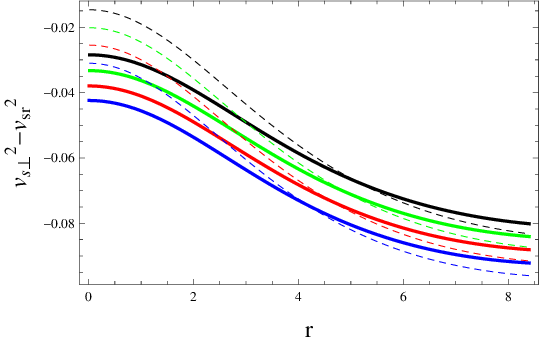,width=0.4\linewidth} \caption{Stability
criteria for $\eta=0.1$ (solid) and $0.3$ (dotted) corresponding to
model II.}
\end{figure}

\subsection{Model III}

Here, we assume a linear equation that connects different components
of $\mathfrak{D}$-sector given as follows \cite{42babc}
\begin{equation}\label{g59}
\mathfrak{D}_{1}^{1}(r)=\tau_1\mathfrak{D}_{0}^{0}(r)+\tau_2,
\end{equation}
where $\tau_1$ and $\tau_2$ are real-valued constants whose
different values can be taken to check how the corresponding
solution is influenced by these unknowns. Making use of
Eqs.\eqref{g22} and \eqref{g23} in the above linear equation yields
\begin{equation}\label{g60}
\mathrm{T}(r)\left(\frac{\nu_1'}{r}+\frac{1}{r^2}-\frac{\tau_1}{r^2}\right)-\frac{\tau_1\mathrm{T}'(r)}{r}-8\pi\tau_2=0.
\end{equation}
One can involve $\mathfrak{D}_{2}^{2}(r)$ component in
Eq.\eqref{g59}, however, its presence will make the above equation
more complicated due to the appearance of second order derivatives
of the metric potentials. Joining the metric ansatz \eqref{g33} and
\eqref{g34} with \eqref{g60}, we get
\begin{align}\nonumber
&\mathrm{T}(r)\bigg\{\frac{6 \mathrm{C}_3 \big(\sqrt{\mathrm{C}_3
r^2+2-\mathrm{C}_3^2 r^4}+\mathrm{C}_2-2 \mathrm{C}_2 \mathrm{C}_3
r^2\big)}{\big(\mathrm{C}_3 r^2+1\big) \sqrt{\mathrm{C}_3
r^2+2-\mathrm{C}_3^2 r^4}-\mathrm{C}_2 \big(\mathrm{C}_3 r^2-2\big)
\big(2\mathrm{C}_3r^2+5\big)}\\\label{g61}
&+\frac{1-\tau_1}{r^2}\bigg\}-\frac{\tau_1\mathrm{T}'(r)}{r}-8\pi\tau_2=0.
\end{align}
This is first order differential equation in $\mathrm{T}(r)$ whose
analytical solution is not possible due to the appearance of a term
in the square root. Therefore, we employ a numerical integration
with an initial condition $\mathrm{T}(0)=0$ to obtain the
deformation function for $\tau_1=1.6$ and $\tau_2=-0.001$.
\begin{table}
\scriptsize \centering \caption{Values of central density, surface
density, central pressure, surface compactness and surface redshift
corresponding to $\eta=0.1$ for Model III.} \label{Table6}
\vspace{+0.07in} \setlength{\tabcolsep}{0.95em}
\begin{tabular}{ccccccc}
\hline\hline $\nu_3$ & 0 & 0.25 & 0.5 & 0.75
\\\hline $\mu_c~(gm/cm^3)$ & 9.5535$\times$10$^{14}$ & 9.2779$\times$10$^{14}$ & 9.0051$\times$10$^{14}$ &
8.7602$\times$10$^{14}$
\\\hline $\mu_s~(gm/cm^3)$ & 5.4691$\times$10$^{14}$ & 5.3166$\times$10$^{14}$ & 5.1346$\times$10$^{14}$ &
4.9513$\times$10$^{14}$
\\\hline $P_c ~(dyne/cm^2)$ & 1.1115$\times$10$^{35}$ & 1.0837$\times$10$^{35}$ & 1.0519$\times$10$^{35}$ &
1.0321$\times$10$^{35}$
\\\hline $\zeta_s$ & 0.151 & 0.147 & 0.143 & 0.139
\\\hline
$z_s$ & 0.199 & 0.191 & 0.184 & 0.176\\
\hline\hline
\end{tabular}
\end{table}
\begin{table}
\scriptsize \centering \caption{Values of central density, surface
density, central pressure, surface compactness and surface redshift
corresponding to $\eta=0.3$ for Model III.} \label{Table7}
\vspace{+0.07in} \setlength{\tabcolsep}{0.95em}
\begin{tabular}{ccccccc}
\hline\hline $\nu_3$ & 0 & 0.25 & 0.5 & 0.75
\\\hline $\mu_c~(gm/cm^3)$ & 8.9127$\times$10$^{14}$ & 8.6384$\times$10$^{14}$ & 8.3642$\times$10$^{14}$ &
8.1194$\times$10$^{14}$
\\\hline $\mu_s~(gm/cm^3)$ & 4.7988$\times$10$^{14}$ & 4.6169$\times$10$^{14}$ & 4.4336$\times$10$^{14}$ &
4.2503$\times$10$^{14}$
\\\hline $P_c ~(dyne/cm^2)$ & 1.1713$\times$10$^{35}$ & 1.1314$\times$10$^{35}$ & 1.1113$\times$10$^{35}$ &
1.0839$\times$10$^{35}$
\\\hline $\zeta_s$ & 0.125 & 0.121 & 0.117 & 0.113
\\\hline
$z_s$ & 0.157 & 0.151 & 0.143 & 0.136\\
\hline\hline
\end{tabular}
\end{table}

Figure \textbf{12} shows increasing trend outwards and in an inverse
relation with the model parameter. We utilize Eqs.\eqref{g13} and
\eqref{g14} to construct the corresponding matter triplet
$(\tilde{\mu},\tilde{P}_r,\tilde{P}_\bot)$ and anisotropy. The
variation of these variables with respect to $r$, $\eta$ and $\nu_3$
is shown in Figure \textbf{13}, indicating an acceptable profile
(see Tables \textbf{6} and \textbf{7} for numerical values). They
show the same behavior as we have found for the first and second
models. Figure \textbf{14} presents the plots of the mass function,
redshift and compactness that are consistent with their acceptable
limits. Further, this solution produces less massive interior in
comparison with models I and II. The variation of the energy
conditions $\tilde{\mu}-\tilde{P}_r \geq 0$ and
$\tilde{\mu}-\tilde{P}_\bot \geq 0$ is shown in Figure \textbf{15},
demonstrating that our model III is viable. Both criteria of
stability (Figure \textbf{16}) indicate that the developed model is
stable for every parametric choice.
\begin{figure}\center
\epsfig{file=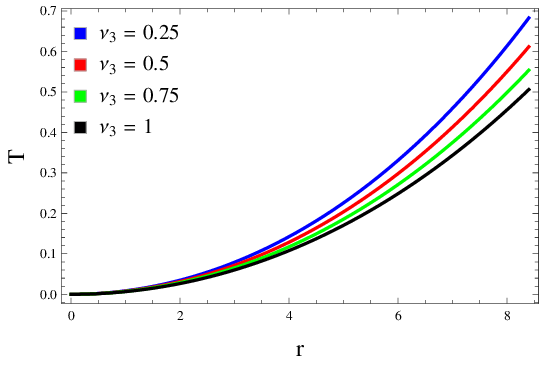,width=0.4\linewidth}
\caption{Deformation function corresponding to model III.}
\end{figure}
\begin{figure}\center
\epsfig{file=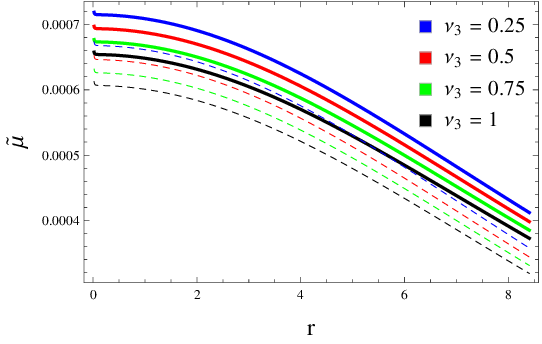,width=0.4\linewidth}\epsfig{file=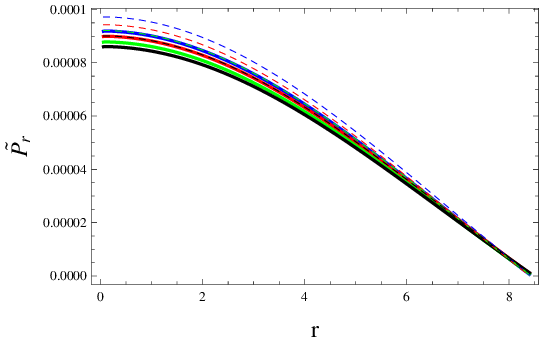,width=0.4\linewidth}
\epsfig{file=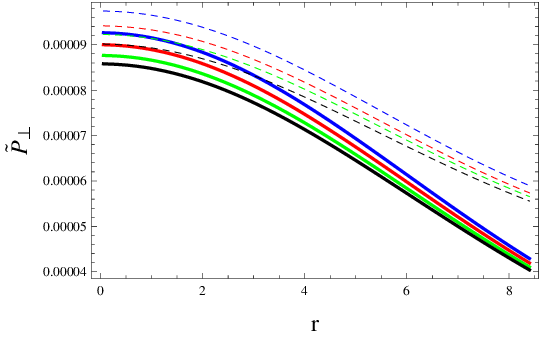,width=0.4\linewidth}\epsfig{file=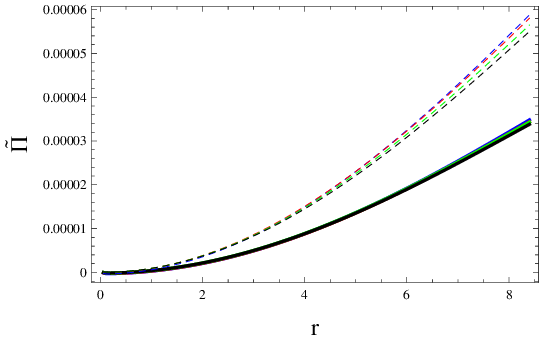,width=0.4\linewidth}
\caption{Physical variables and anisotropy for $\eta=0.1$ (solid)
and $0.3$ (dotted) corresponding to model III.}
\end{figure}
\begin{figure}\center
\epsfig{file=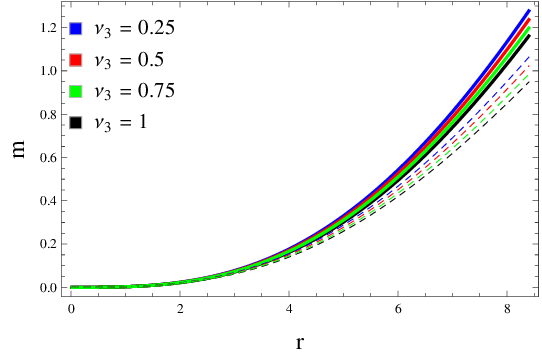,width=0.4\linewidth}\epsfig{file=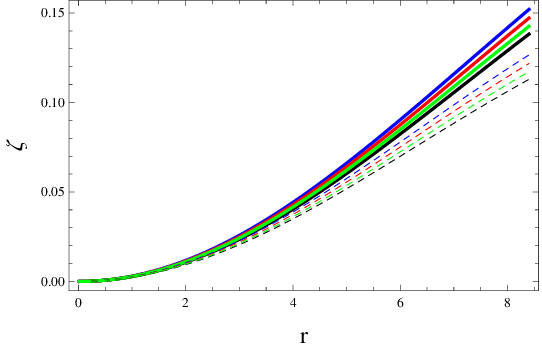,width=0.4\linewidth}
\epsfig{file=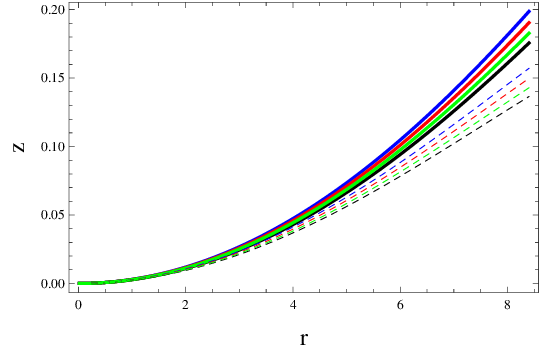,width=0.4\linewidth} \caption{Different
parameters for $\eta=0.1$ (solid) and $0.3$ (dotted) corresponding
to model III.}
\end{figure}
\begin{figure}\center
\epsfig{file=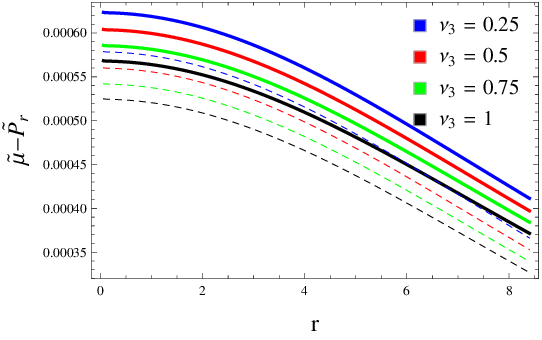,width=0.4\linewidth}\epsfig{file=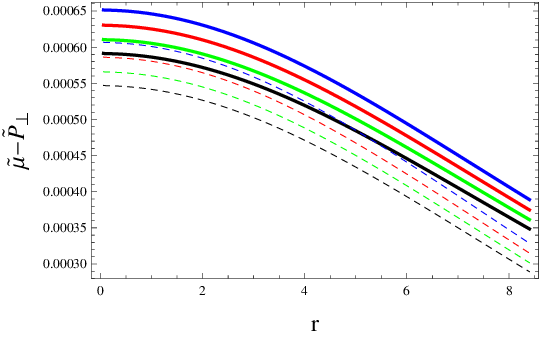,width=0.4\linewidth}
\caption{Dominant energy bounds for $\eta=0.1$ (solid) and $0.3$
(dotted) corresponding to model III.}
\end{figure}
\begin{figure}\center
\epsfig{file=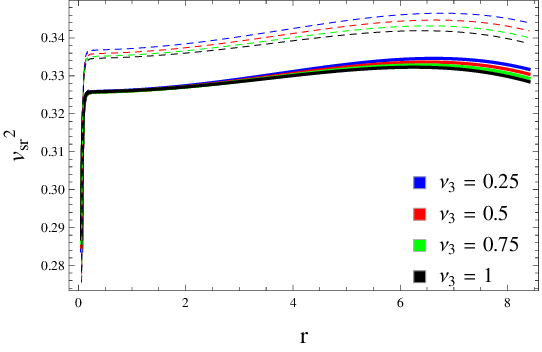,width=0.4\linewidth}\epsfig{file=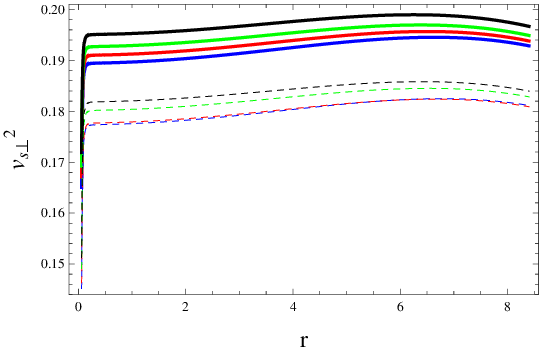,width=0.4\linewidth}
\epsfig{file=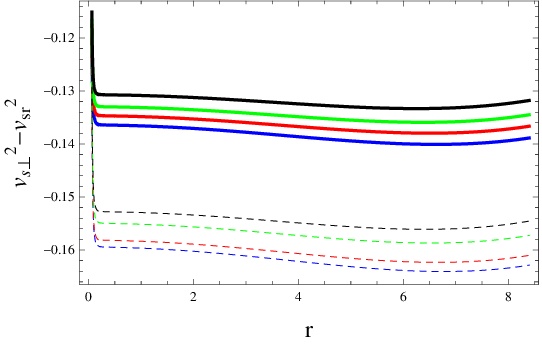,width=0.4\linewidth} \caption{Stability
criteria for $\eta=0.1$ (solid) and $0.3$ (dotted) corresponding to
model III.}
\end{figure}

\section{Conclusions}

In this paper, we have formulated different anisotropic extensions
of the known isotropic solution through gravitational decoupling for
the model $f(\mathbb{R},\mathbb{T})=\mathbb{R}+2\nu_3\mathbb{T}$.
Initially, we have assumed that a sphere is filled with an isotropic
fluid that becomes anisotropic after inserting an additional source
at the action level. The modified form of the action \eqref{g1} has
then triggered the field equations involving the effects of both
seed isotropic as well as additional sources and modified theory. We
have then split these equations into two sets through MGD strategy.
We have dealt with the unknowns of the first set by choosing
Buchdahl anstaz given by
\begin{align}\nonumber
e^{\nu_1(r)}&=\mathrm{C}_{1}\left[\big(1+\mathrm{C}_{3}r^2\big)^{\frac{3}{2}}+\mathrm{C}_{2}\big(5+2\mathrm{C}_{3}r^2\big)
\sqrt{2-\mathrm{C}_{3}r^2}\right]^2,\\\nonumber
e^{\nu_2(r)}&=\frac{2\big(1+\mathrm{C}_{3}r^2\big)}{2-\mathrm{C}_{3}r^2},
\end{align}
and calculated the triplet
$(\mathrm{C}_{1},\mathrm{C}_{2},\mathrm{C}_{3})$ through junction
conditions at $\Sigma:~r=\mathrm{R}$. Moreover, the
$\mathfrak{D}$-sector \eqref{g21}-\eqref{g23} comprised four
unknowns that needed to be determined. Therefore, we have imposed
different constraints on $\mathfrak{D}_{\sigma\omega}$, resulting in
three different solutions. We have then added these solutions
corresponding to both fluid sources through the parameter $\eta$ and
obtained new anisotropic extensions.

We have discussed acceptability conditions such that the resulting
solution must be in agrement with them. Such physical
characteristics of all the developed models have been interpreted
graphically for $\nu_3=0.25,0.5,0.75,1$ and $\eta=0.1,0.3$. We have
observed an acceptable behavior of the matter determinants and
anisotropy in each case for all parametric values. The numerical
values of the mass function indicates that the second model produces
massive interior in comparison to the others. The redshift and
compactness factors have also been shown compatible with the
experimental data. The dominant energy bounds are satisfied in the
region $0<r<\mathrm{R}$, ensuring the viability of the obtained
solutions. Finally, we have noticed that the cracking occurs only in
the first model, hence, models II and III are stable.

Morales and Tello-Ortiz \cite{bb} extended the Durgapal's fifth
model to the anisotropic domain through the MGD approach and
obtained the more dense structure in comparison with our developed
solutions. Andrade and Contreras \cite{cc} used Tolman IV,
Heintzmann IIa and Durgapal IV as seed solutions, and proposed
different anisotropic extensions by using the definition of the
complexity factor. The anisotropy was also discussed in the interior
of SMC X-1 and Cen X-3 stars from which we found this factor to be
much greater as compared to that obtained in the current setup.
Maurya et al. \cite{dd} investigated the possible existence of
compact stars influenced by the electromagnetic field in the
framework of Gauss-Bonnet gravity and determined the exact solutions
to the corresponding field equations in contrast with our work. It
is important to stress here that our models I, II and III are
consistent with \cite{34}, \cite{25ac} and \cite{42bc},
respectively. All our results reduce to $\mathbb{GR}$ for
$\nu_3=0$.\\\\
\textbf{Data Availability Statement:} This manuscript has no
associated data.


\vspace{0.25cm}

\begin{thebibliography}{10}
\bibitem{2} Capozziello, S. et al.: Class. Quantum Grav. \textbf{25}(2008)
085004.

\bibitem{3} Nojiri, S. et al.: Phys. Lett. B \textbf{681}(2009)74.

\bibitem{2a} de Felice, A. and Tsujikawa, S.: Living Rev.
Relativ. \textbf{13}(2010)3.

\bibitem{4} Nojiri, S. and Odintsov, S.D.: Phys.
Rep. \textbf{505}(2011)59.

\bibitem{9} Sharif, M. and Kausar, H.R.: J. Cosmol. Astropart. Phys.
\textbf{07}(2011)022.

\bibitem{1n} Astashenok, A.V., Capozziello, S., Odintsov, S.D. and Oikonomou, V.K.: Phys. Lett. B \textbf{816}(2021)136222.

\bibitem{1o} Astashenok, A.V., Capozziello, S. and Odintsov, S.D.: J. Cosmol. Astropart. Phys. \textbf{12}(2013)040.

\bibitem{1p} Astashenok, A.V., Capozziello, S. and Odintsov, S.D.: Astrophys. Space Sci. \textbf{355}(2015)333.

\bibitem{1q} Astashenok, A.V., Odintsov, S.D. and De la Cruz-Dombriz, A.: Class. Quantum Grav. \textbf{34}(2017)205008.

\bibitem{1r} Astashenok, A.V., Capozziello, S., Odintsov, S.D. and Oikonomou, V.K.: Phys. Lett. B \textbf{811}(2020)135910.

\bibitem{1s} Astashenok, A.V., Capozziello, S., Odintsov, S.D. and Oikonomou, V.K.: EPL \textbf{136}(2022)59001.

\bibitem{1t} Astashenok, A.V., Odintsov, S.D. and Oikonomou, V.K.: Phys. Rev. D \textbf{106}(2022)124010.

\bibitem{10} Bertolami, O. et al.: Phys. Rev. D \textbf{75}(2007)104016.

\bibitem{11} Naseer, T., Sharif, M., Fatima, A. and Manzoor, S.: Chin. J. Phys. \textbf{86}(2023)350.

\bibitem{20} Harko, T. et al.: Phys. Rev. D \textbf{84}(2011)024020.

\bibitem{22} Deng, X.M. and Xie, Y.: Int. J. Theor. Phys.
\textbf{54}(2015)1739.

\bibitem{22a} Houndjo, M.J.S.: Int. J. Mod. Phys. D
\textbf{21}(2012)1250003.

\bibitem{22b} Das, A. et al.: Phys. Rev. D \textbf{95}(2017)124011.

\bibitem{25ae} Singh, K.N. et al.: Phys. Dark Universe
\textbf{30}(2020)100620.

\bibitem{25af} Maurya, S.K.: Phys. Dark Universe \textbf{30}(2020)100640.

\bibitem{25b} Rej, P., Bhar, Piyali. and Govender, M.: Eur. Phys. J. C
\textbf{81}(2021)316.

\bibitem{25afa} Kaur, S., Maurya, S.K., Shukla, S. and Nag, R.:
Chin. J. Phys. \textbf{77}(2022)2854.

\bibitem{25ad} Sharif, M. and Naseer, T.: Eur. Phys. J. Plus
\textbf{137}(2022)1304.

\bibitem{25ada} Sharif, M. and Naseer, T.: Phys. Scr.
\textbf{98}(2023)115012.

\bibitem{25adb} Sharif, M. and Naseer, T.: Ann. Phys.
\textbf{459}(2023)169527.

\bibitem{1i} Zubair, M., Waheed, S. and Ahmad, Y.: Eur. Phys. J. C
\textbf{76}(2016)444.

\bibitem{1j} Das, A., Rahaman, F., Guha, B.K. and Ray, S.: Eur.
Phys. J. C \textbf{76}(2016)654.

\bibitem{1k} Moraes, P.H.R.S., Correa, R.A.C. and Lobato, R.V.: J. Cosmol. Astropart. Phys.
\textbf{07}(2017)029.

\bibitem{1l} Zaregonbadi, R. et al.: Phys. Rev. D \textbf{94}(2016)084052.

\bibitem{25ag} Buniy, R.V., Berera, A. and Kephart, T.W.: Phys. Rev. D
\textbf{73}(2006)063529.

\bibitem{25aga} Saadeh, D., Feeney, S.M., Pontzen, A., Peiris, H.V. and McEwen, J.D.: Phys.
Rev. Lett. \textbf{117}(2016)131302.

\bibitem{25ah} Mishra, B., Ray, P.P. and
Myrzakulov, R.: Eur. Phys. J. C \textbf{79}(2019)34.

\bibitem{25ai} Sokolov, A.I.: Sov. Phys.-JETP \textbf{52}(1980)575.

\bibitem{25aj} Sawyer, R.F.: Phys. Rev. Lett. \textbf{29}(1972)382.

\bibitem{25ak} Weber, F.: J. Phys. G: Nucl. Part. Phys. \textbf{25}(1999)R195.

\bibitem{25al} Schunck, F.E. and Mielke, E.W.: Class. Quantum Grav.
\textbf{20}(2003)R301.

\bibitem{25ala} Rahaman, F., Ray, S., Jafry, A.K. and Chakraborty, K.: Phys. Rev. D
\textbf{82}(2010)104055.

\bibitem{25am} Maurya, S.K., Mishra, B., Ray, S. and Nag, R.: Chin. Phys. C
\textbf{46}(2022)105105.

\bibitem{25an} Migkas, K. and Reiprich, T.H.: Astron. Astrophys. \textbf{611}(2018)A50.

\bibitem{25ao} Migkas, K., Schellenberger, G., Reiprich, T.H., Pacaud, F., Ramos-Ceja, M.E. and Lovisari, L.:
Astron. Astrophys. \textbf{636}(2020)A15.

\bibitem{29} Ovalle, J.: Mod. Phys. Lett. A \textbf{23}(2008)3247.

\bibitem{30} Ovalle, J. and Linares, F.: Phys. Rev. D \textbf{88}(2013)104026.

\bibitem{31} Casadio, R., Ovalle, J. and Da Rocha, R.: Class. Quantum Grav. \textbf{32}(2015)215020.

\bibitem{33} Ovalle, J. et al.: Eur. Phys. J. C \textbf{78}(2018)960.

\bibitem{34} Sharif, M. and Sadiq, S.: Eur. Phys. J. C \textbf{78}(2018)410.

\bibitem{35} Sharif, M. and Waseem, A.: Ann. Phys. \textbf{405}(2019)14.

\bibitem{36} Gabbanelli, L., Rinc{\'o}n, {\'A}. and Rubio, C.: Eur. Phys. J. C \textbf{78}(2018)370.

\bibitem{36a} Estrada, M. and Tello-Ortiz, F.: Eur. Phys. J. Plus \textbf{133}(2018)453.

\bibitem{37a} Hensh, S. and Stuchl{\'\i}k, Z.: Eur. Phys. J. C \textbf{79}(2019)834.

\bibitem{37b} Sharif, M. and Ama-Tul-Mughani, Q.: Mod. Phys. Lett. A
\textbf{35}(2020)2050091.

\bibitem{37f} Sharif, M. and Naseer, T.: Chin. J. Phys. \textbf{73}(2021)179.

\bibitem{37fa} Sharif, M. and Naseer, T.: Int. J. Mod. Phys. D \textbf{31}(2022)2240017.

\bibitem{37fb} Sharif, M. and Naseer, T.: Indian J. Phys. \textbf{96}(2022)4373.

\bibitem{37fc} Naseer, T. and Sharif, M.: Universe \textbf{8}(2022)62.

\bibitem{38} Ashmita, Sarkar, P. and Das, P.K.: Int. J. Mod. Phys. D
\textbf{31}(2022)2250120.

\bibitem{39} Kaur, S., Maurya, S.K., Shukla, S. and Nag, R.: Chin.
J. Phys. \textbf{77}(2022)2854.

\bibitem{40} Sharif, M. and Naseer, T.: Class. Quantum Grav. \textbf{40}(2023)035009.

\bibitem{41} Naseer, T. and Sharif, M.: Fortschr. Phys. \textbf{71}(2023)2300004.

\bibitem{aa} Maurya, S.K. and Tello-Ortiz, F.: Phys. Dark Universe
\textbf{27}(2020)100442.

\bibitem{aaa} Maurya, S.K., Mishra, B., Ray, S. and Nag, R.: Chin. Phys. C
\textbf{46}(2022)105105.

\bibitem{42a} Buchdahl, H.A.: Phys. Rev. \textbf{116}(1959)1027.

\bibitem{42} Vaidya, P.C. and Tikekar, R.: J. Astrophys. Astro.
\textbf{3}(1982)325.

\bibitem{58} Kumar, J., Prasad, A.K., Maurya, S.K. and Banerjee, A.: Eur. Phys.
J. C \textbf{78}(2018)540.

\bibitem{63} Sharma, R., Karmakar, S. and Mukherjee, S.: Int. J. Mod. Phys. D
\textbf{15}(2006)405.

\bibitem{a} Maurya, S.K. et al.: Phys. Rev. D
\textbf{99}(2019)044029.

\bibitem{b} Singh, K.N., Pant, N. and Pradhan, N.: Astrophys. Space Sci. \textbf{361}(2016)173.

\bibitem{c} Maurya, S.K., Banerjee, A. and Tello-Ortiz, F.: Phys.
Dark Universe \textbf{27}(2020)100438.

\bibitem{d} Maurya, S.K., Singh, K.N., Govender, M. and Ray, S.: Fortschr. Phys.
\textbf{71}(2023)2300023.

\bibitem{42aa} Gangopadhyay, T., Ray, S., Xiang-Dong, L., Jishnu, D. and Mira, D.: Mon. Not. R. Astron.
Soc. \textbf{431}(2013)3216.

\bibitem{ab} Delgaty, M.S.R. and Lake, K.: Comput. Phys. Commun. \textbf{115}(1998)395.

\bibitem{ac} Ivanov, B.V.: Eur. Phys. J. C \textbf{77}(2017)738.

\bibitem{ad} Sharif, M. and Naseer, T.: Phys. Dark Universe \textbf{42}(2023)101324.

\bibitem{ae} Sharif, M. and Naseer, T.: Chin. J. Phys.
\textbf{86}(2023)596.

\bibitem{af} Sharif, M. and Naseer, T.: Gen. Relativ. Gravit.
\textbf{55}(2023)87.

\bibitem{aza} Naseer, T. et al.: Mod. Phys. Lett. A \textbf{39}(2024)2450048

\bibitem{azb} Feng, Y. et al.: Chin. J. Phys. \textbf{90}(2024)372.

\bibitem{azc} Feng, Y. et al.: Phys. Scr. \textbf{99}(2024)085034.

\bibitem{azd} Naseer, T. and Sharif, M.: Chin. J. Phys. \textbf{88}(2024)10.

\bibitem{42b} Ivanov, B.V.: Phys. Rev. D \textbf{65}(2002)104011.

\bibitem{42bb} Abreu, H., Hernandez, H. and Nunez, L.A.: Class. Quantum Grav.
\textbf{24}(2007)4631.

\bibitem{42ba} Herrera, L.: Phys. Lett. A \textbf{165}(1992)206.

\bibitem{42baba} Heras, C.L. and Le{\'o}n, P.: Fortschr. Phys. \textbf{66}(2018)1800036.

\bibitem{42babb} Graterol, R.P.: Eur. Phys. J. Plus
\textbf{133}(2018)244.

\bibitem{42babc} Contreras, E. and Bargue{\~n}o, P.: Class. Quantum
Grav. \textbf{36}(2019)215009.

\bibitem{bb} Morales, E. and Tello-Ortiz, F.: Eur. Phys. J. C \textbf{78}(2018)841.

\bibitem{cc} Andrade, J. and Contreras, E.: Eur. Phys. J. C \textbf{81}(2021)889.

\bibitem{dd} Maurya, S.K. et al.: Eur. Phys. J. C
\textbf{82}(2022)552.

\bibitem{25ac} Azmat, H. and Zubair M.: Eur. Phys. J. Plus
\textbf{136}(2021)112.

\bibitem{42bc} Al Hadhrami, M. et al.: Pramana \textbf{97}(2022)13.
\end{thebibliography}
\end{document}